\newcommand{\be}{\begin{equation}}\newcommand{\ee}{\end{equation}}
\newcommand{\bea}{\begin{eqnarray}}\newcommand{\eea}{\end{eqnarray}}
\newcommand{\nn}{\nonumber\\[6pt]}
\newcommand{\lb}[1]{\label{#1}}
\newcommand{\p}[1]{(\ref{#1})}

\newcommand{\eps}{\epsilon}

\documentclass[12pt]{article}
\usepackage{amscd,amsmath,amssymb}

\begin{document}

\thispagestyle{empty}
\vspace{2cm}
\begin{flushright}
ITP-UH-11/07 
\\[1cm]
\end{flushright}
\begin{center}
{\Large\bf Hierarchy of N=8 Mechanics Models}
\end{center}
\vspace{1cm}

\begin{center}
{\large\bf
Evgeny Ivanov${\,}^{a}$, \
Olaf Lechtenfeld${\,}^{b}$ \ and \
Anton Sutulin${\,}^{a}$ }
\end{center}

\begin{center}

\vspace{0.2cm}
${}^a$ {\it
Bogoliubov  Laboratory of Theoretical Physics, JINR, 141980 Dubna, Russia}

\vspace{0.2cm}
${}^b$ {\it
Institut f\"ur Theoretische Physik, Leibniz Universit\"at Hannover,} \\
{\it Appelstra\ss{}e 2, 30167 Hannover, Germany}
\vspace{0.2cm}

{\tt eivanov, sutulin@theor.jinr.ru,  lechtenf@itp.uni-hannover.de}
\end{center}
\vspace{2cm}

\begin{abstract}
\noindent
Using the $N{=}4$ superspace approach in one dimension (time),
we construct general $N{=}8$ supersymmetric mechanics actions for the
multiplets ${\bf(b,8,8{-}b)}$ classified in {\tt hep-th/0406015},
with the main focus on the previously unexplored cases of
${\bf(8,8,0)}$, ${\bf(7,8,1)}$ and ${\bf(6,8,2)}\,$, as well
as on ${\bf(5,8,3)}$ for completeness.
$N{=}8$ supersymmetry of the action amounts to a harmonicity condition for
the Lagrangian with respect to its superfield arguments.
We derive the generic off-shell component action for the ``root'' multiplet
${\bf(8,8,0)}$, prove that the actions for all other multiplets follow from it
through automorphic dualities and argue that this hierarchical structure
is universal.
The bosonic target geometry in all cases is conformally flat,
with a unique scalar potential (except for the root multiplet).
We show that the $N{=}4$ superfield constraints respect the full $R$-symmetry
and find the explicit realization of its quotient over the manifest $R$-symmetry
on superfields and component fields. Several $R$-symmetric $N{=}4$ superfield Lagrangians
with $N{=}8$ supersymmetry are either newly found or reproduced by a simple universal method.
\end{abstract}
\vfill

\newpage
\setcounter{page}{1}
\section{Introduction}
Particle models with extended supersymmetry supply nontrivial examples of
supersymmetric mechanical systems and of supersymmetric quantum mechanics
(see e.g.~\cite{Lima,BMSV,AnS}).
In many cases they are related by dimensional reduction
to some higher-dimensional supersymmetric field theory,
e.g.~$N{=}4$ super Yang-Mills theory (see e.g.~\cite{BLY}),
and as such provide a simplified setting for studying salient features
of these theories. Other models of this kind represent supersymmetric
extensions of certain intrinsically one-dimensional systems, like
conformal mechanics~\cite{AP}-\cite{BIKL2},~\cite{BMSV},
or Calogero-Moser integrable models~\cite{Freed}-\cite{GL+}.
{}From the mathematical point of view, extended $d{=}1$ supersymmetry,
as compared with its higher-dimensional counterpart, exhibits rather
unusual features, such as the so-called automorphic duality between
multiplets with the same number of fermions but with different distributions
of the bosons to the physical and the auxiliary sector~\cite{GR,GR1,PT}.
Furthermore, the $d{=}1$ systems with extended supersymmetry display
special types of bosonic target geometries which are not encountered in
higher-dimensional supersymmetric field theories~\cite{Geom,GPS}.

Until now, most exhaustively studied have been the mechanics models
with $2\leq N \leq 4$ supersymmetries
(see e.g.\ the review~\cite{Lima} and refs.\ therein).
Clearly, the higher is $N$, the more restrictions on the dynamics and
target geometry of the corresponding mechanics models are expected. From
this point of view, models with $N{=}8$ supersymmetry are of obvious interest.
The corresponding supersymmetry algebra can be obtained by direct dimensional
reduction from the four-dimensional $N{=}2$ Poincar\'e superalgebra.
Therefore, some of $N{=}8$ mechanics models are dimensional reductions of
$N{=}2$ super Yang-Mills theory or models of self-interacting $N{=}2$
hypermultiplets in four dimensions. The $d{=}1$ reduction of the gauge
multiplet can yield either one of the off-shell $N{=}8$ multiplets
${\bf(5,8,3)}$ and ${\bf(2,8,6)}$
(in the classification of refs.~\cite{BIKL2,ABC}\footnote{
The symbol ${\bf(b,f,a)}$ denotes the off-shell $d{=}1$ multiplet
with ${\bf f}$ physical fermionic, ${\bf b}$ physical bosonic and
${\bf a}$ auxiliary bosonic fields, with ${\bf f}={\bf b}+{\bf a}$.}),
depending on whether one performs the reduction on the level of the
gauge potential superfield \cite{DE}-\cite{ISm} or the superfield strength
\cite{BKN}-\cite{BKS1}.
The $d{=}1$ reduction of the generic hypermultiplet model
(which exists off shell only in harmonic superspace~\cite{harm,book})
produces a general one-dimensional sigma model with hyperk\"ahler target
geometry and an infinite number of auxiliary fields.
Finally, the $d{=}1$ reduction of the general off-shell sigma model for
twisted $N{=}(4,4)$ supermultiplets in two dimensions leads to the most
general model of the $N{=}8$ multiplet ${\bf(4,8,4)}$~\cite{BKSu,BIS}.

It should be pointed out, however, that by dimensional reduction one does not
exhaust the full set of off-shell $d{=}1$ supermultiplets and corresponding
supersymmetric mechanics systems. For instance, in the case of
$N{=}4$ and $N{=}8$ supersymmetry there exist off-shell multiplets
with field contents ${\bf(4,4,0)}$ and ${\bf(8,8,0)}$, respectively
(so-called ``root''~\cite{Faux,root} or ``extreme'' multiplets). These cannot
be recovered by dimensional reduction because the suitable $d{>}1$ off-shell
multiplets always include auxiliary fields.
Therefore, in order to compile a complete list of $d{=}1$ supermultiplets
and relevant supersymmetric mechanics models, we better not refer to
dimensional reduction but proceed solely from the $d{=}1$ supersymmetry
algebra under consideration. This strategy was pursued in the papers
\cite{ikl1, BIKL2, ABC} co-authored by two of the present authors
(E.I.\ \& O.L.).
Using superfield techniques, we derived the full set
$\{{\bf(b,4,4{-}b)}\,|\,b=0,\dots,4\}$ of off-shell $N{=}4$
multiplets with four physical fermions~\cite{ikl1} and the full set
$\{{\bf(b,8,8{-}b)}\,|\,b=0,\dots,8\}$ of linear off-shell $N{=}8$ multiplets
with eight physical fermions~\cite{ABC}.
These multiplets can be defined either by manifestly $N{=}8$ supersymmetric
constraints in $N{=}8, d{=}1$ superspace or by equivalent sets of constraints
in $N{=}4, d{=}1$ superspace, with one $N{=}4$ supersymmetry being manifest
and a second one hidden.

In \cite{ABC}, by exploiting the $N{=}4$ superfield formalism,
the free actions for all mentioned $N{=}8$ multiplets were constructed,
and particular examples of nontrivial $N{=}8$ invariant actions were presented.
Later on, some of these multiplets (as well as their nonlinear cousins) and
associated $N{=}8$ mechanics were studied in more detail in
\cite{BKN,BKNS1,BKS1,BKSu} and \cite{BBKS}-\cite{BurKS}.
The general sigma-model type action for the multiplet
${\bf(5,8,3)}$, both in $N{=}4$ superfield formulation and in $N{=}8$
harmonic superfield formulation, had been constructed earlier
in~\cite{DE}-\cite{ISm}.
Superconformally invariant actions for the multiplets ${\bf(3,8,5)}$ and
${\bf(5,8,3)}$, in $N{=}4$ superfield formulation, were given in~\cite{BIKL2}.
Manifestly $N{=}8$ supersymmetric general actions in terms of constrained
$N{=}8$ superfields were given for the multiplet ${\bf(4,8,4)}$
(in bi-harmonic $N{=}8$ superspace)~\cite{BIS} and ${\bf(2,8,6)}$
(in conventional $N{=}8, d{=}1$ superspace)~\cite{BKN,BKNS1}.

As for the remaining $N{=}8$ multiplets, namely ${\bf(0,8,8)}$, ${\bf(1,8,7)}$,
${\bf(6,8,2)}$, ${\bf(7,8,1)}$ and ${\bf(8,8,0)}$, no detailed analysis of
the corresponding models or construction of generic invariant actions have
been carried out yet.
One reason is that no convenient $N{=}8$ superfield formalism is known for
describing these multiplets. Although their defining off-shell constraints
can be written in conventional $N{=}8$ superspace~\cite{ABC},
these constraints cannot be cast (at least not in a simple way) in the form
of Grassmann analyticity conditions and then solved in terms of analytic
superfields living on $N{=}8$ superspaces of smaller Grassmann dimension.
On the other hand, by dimensional reasoning, the sigma-model-type actions
of these multiplets cannot be written as integrals of the relevant constrained
$N{=}8$ superfields over the full $N{=}8$ superspace. Typically, one may
overcome this difficulty by solving the constraints through negative-dimension
prepotentials; the appearance of new non-geometric gauge invariances in
such a prepotential formulation renders it impractical however.

Another, much more feasible way around is to employ the off-shell $N{=}4$
superfield approach. In this paper, we use it to find the general $N{=}8$
supersymmetric actions for the so far unexplored multiplets
${\bf(8,8,0)}$, ${\bf(7,8,1)}$ and ${\bf(6,8,2)}$. For completeness,
we also study the ${\bf(5,8,3)}$ case within the same setting, for the
$N{=}4$ superfield splitting ${\bf(5,8,3)}={\bf(4,4,0)}\oplus{\bf(1,4,3)}$
which was not fully discussed in the literature before.
We derive general conditions for the existence of the hidden $N{=}4$
supersymmetry in the superfield actions of the constituent $N{=}4$ superfields
(and hence for the full $N{=}8$ supersymmetry). We show that in all cases
it is the harmonicity of the Lagrangian with respect to its superfield
arguments which guarantees this second $N{=}4$ supersymmetry.
For all cases considered we derive the general bosonic action and, in addition,
the full general component action for the ``root'' multiplet ${\bf(8,8,0)}$.
We argue that the generic sigma-model-type off-shell component action of
{\it each\/} $N{=}8$ model ${\bf(b,8,8{-}b)}$ can be obtained from this
${\bf(8,8,0)}$ ``parent'' action via automorphic duality and demonstrate this
property for~${\bf b}\ge5$.  This hierarchical structure of the $N{=}8$ mechanics
models generalizes the $N{=}4$ hierarchy established in~\cite{root}.

The target bosonic geometry is always conformally flat, with the conformal
factor obeying the appropriate ${\bf b}$-dimensional harmonicity
condition (with ${\bf b}\ge5$). We also present explicit examples of actions
enjoying invariance under the corresponding $R$-symmetry. It is shown that
the $N{=}4$ superfield constraints respect the full
SO$({\bf b})\times $SO$({\bf 8{-}b})$ symmetry, which becomes the maximal
$R$-symmetry group SO(8) of the $N{=}8$, $d{=}1$ Poincar\'e superalgebra
in the extreme case of the multiplet ${\bf(8,8,0)}$.
We display the realization of the hidden parts of these $R$-symmetries both on the
$N{=}4$ superfields and on the component fields and give the explicit form
of the corresponding invariant $N{=}4$ actions and their bosonic cores.
In all cases, except for ${\bf(8,8,0)}$, an $N{=}8$ potential term exists
and is written down.

\section{The (8,8,0) mechanics}
\subsection{N=8 invariant action in N=4 superspace}
We will use the $N{=}4$ superspace parametrized by $z=(t, \theta^{ia})$, with the spinor
covariant derivatives $D^{ia} = \frac{\partial}{\partial \theta_{ia}} + i\theta^{ia}\partial_t$
satisfying the anticommutation relation
\be
\{D^{ia}, D^{kb}\} = 2i\epsilon^{ik}\epsilon^{ab}\partial_t\,. \lb{N4acomm}
\ee
The multiplet ${\bf (8,8,0)}$ is presented by two different off-shell $N{=}4$
supermultiplets ${\bf (4,4,0)}$ described by the quartet superfields $\phi^{i\alpha}$
and $q^{a A}$, $\alpha$ and $A$ being doublet indices of two extra SU(2) groups
of the Pauli-G\"ursey type commuting with the $N{=}4$ superalgebra generators (and with the
covariant derivatives as well) \cite{ABC}. These superfields are subjected to the constraints
\be
D^{a(i}\phi^{k)\alpha} =0\,, \quad D^{i(a}q^{b)A} =0\,. \lb{N488constr}
\ee
The hidden $N{=}4$ supersymmetry extending the explicit one to $N{=}8, d{=}1$
supersymmetry is realized by the following transformations mixing $\phi^{i\alpha}$
and $q^{a A}$:
\be
\delta_\eta \phi^{i\alpha} = \frac{1}{2}\eta^{\alpha A}D^{ia} q_{a A}\,, \quad
\delta_\eta q^{a A} = -\frac{1}{2}\eta^{A\alpha} D^{ia}\phi_{i\alpha}\,, \lb{hidd88}
\ee
where $\eta^{A\alpha}$ is the corresponding Grassmann parameter.

The most general sigma-model type action of these two multiplets in $N{=}4$
superspace is given by
\be
S^{\rm gen}_{(8)} = \int dt d^4\theta\, {\cal L}(\phi, q)\,, \lb{2gen}
\ee
where the Grassmann integration measure is defined as
\be
d^4\theta = \frac{1}{4!}\,D^{ai}D^k_a D^b_iD_{kb}\,. \lb{measure}
\ee
To extract the most general $N{=}8$ invariant subclass of the actions \p{2gen},
we should impose on \p{2gen} the condition that it is invariant under \p{hidd88} up to
a total derivative in the integrand
\be
\delta_\eta {\cal L} = \eta_{\alpha A}D_{ai} G^{i\alpha\;aA}(\phi, q)\,. \lb{deltaL1}
\ee
Here $G^{i\alpha\;aA}$ is a function of the involved superfields, arbitrary for the moment.
The $t$-derivative on its own cannot appear in \p{deltaL1}, since it does not
show up in the transformation laws \p{hidd88}. Taking into account that the
constraints \p{N488constr} imply
\be
D^{ia} q^{bA} = \frac{1}{2}\epsilon^{ba} D^{ic} q^A_c \equiv
\frac{1}{2}\epsilon^{ba}\chi^{iA}\,, \quad
D^{ia} \phi^{k\alpha} = \frac{1}{2}\epsilon^{ki} D^{ja} \phi_j^\alpha \equiv
\frac{1}{2}\epsilon^{ki}\chi^{a\alpha}\,, \lb{Corol0}
\ee
and equating the coefficient before the independent spinors $\chi^{iA}$ and
$\chi^{a\alpha}$ in both sides of the condition \p{deltaL1}, we find the
conditions of the hidden $N{=}4$ (and hence full $N{=}8$) supersymmetry of the action \p{2gen}
\be
\mbox{(a)}\;\frac{\partial G^{i\alpha\;aA}}{\partial \phi^{i\beta}}
= -\delta^\alpha_\beta \frac{\partial {\cal L}}{\partial q_{a A}}\,, \quad
\mbox{(b)}\;\frac{\partial G^{i\alpha\;aA}}{\partial q^{a B}}
= \delta^A_B \frac{\partial {\cal L}}{\partial \phi_{i \alpha}}\,. \lb{Cond1}
\ee

As the integrability condition of this system it follows that ${\cal L}$
should satisfy the dimension-8 harmonicity equation
\be
\Delta_{(8)} {\cal L} = (\Delta_{(q)} + \Delta_{(\phi)}){\cal L} = 0\,, \lb{Harm1}
\ee
where
\be
\Delta_{(q)} = \frac{\partial^2}{\partial q^{a A}\,\partial q_{a A}}\,, \quad
\Delta_{(\phi)} = \frac{\partial^2}{\partial \phi^{i\alpha}\,\partial \phi_{i\alpha}}\,.
\ee
One more corollary of \p{Cond1} is
\be
\Delta_{(q)} G^{i\alpha\;aA} = -\Delta_{(\phi)} G^{i\alpha\;aA} =
2\, \frac{\partial^2 {\cal L}}{\partial \phi_{i \alpha}\,\partial q_{a A}}\;\Rightarrow \;
\Delta_{(8)} G^{i\alpha\;aA} = 0\,. \lb{Harm2}
\ee

It is instructive to write \p{Cond1} in a vector form, by the correspondence
$(a A) \leftrightarrow \hat{n}\,, \;
(i \alpha) \leftrightarrow \hat{\mu}$:
\be
\mbox{(a)}\;\partial_{[\hat{\nu}} G_{\hat{\mu]}\;\hat{n}}|_{sd} = 0\,, \;
\partial_{\hat{\mu}} G_{\hat{\mu}\;\hat{n}} = -\partial_{\hat{n}}{\cal L}\,, \quad
\mbox{(b)}\; \partial_{[\hat{m}} G_{\hat{\mu}\;\hat{n}]}|_{sd} = 0\,, \;
\partial_{\hat{n}} G_{\hat{\mu}\;\hat{n}} = \partial_{\hat{\mu}}{\cal L}\,,\lb{Cond11}
\ee
where $|_{sd}$ means self-dual part of the corresponding $R^4$ ``field strength''.
Thus, the $4\times 4$ matrix $G_{\hat{\mu}\;\hat{n}}$ should satisfy two independent
Euclidean self-duality conditions with respect to each vector index, and also its two divergences
should be related to ${\cal L}$ as in \p{Cond11}.

It is worth pointing out that eq. \p{Harm1} is the only integrability condition for ${\cal L}$ which
follows from the general conditions of the invariance of the $N{=}4$ superfield action \p{2gen} under the
extra $N{=}4$ supersymmetry \p{hidd88}. For any ${\cal L}$ satisfying \p{Harm1} eqs. \p{Cond1} and their consequence
\p{Harm2} define the corresponding function $G^{i\alpha\;aA}$ and so ensure the $N{=}8$
supersymmetry of the relevant action. Although the precise form of this function has no practical
value, it can be explicitly found for some particular solutions
of \p{Harm1} (see Subsection 2.2). The fact that \p{Harm1} is the necessary and sufficient condition of the
$N{=}4$ superfield action \p{2gen} to possess $N{=}8$ supersymmetry can be also confirmed by considering the full
off-shell component form of this action (see Subsection 2.4.). This action is parametrized by two arbitrary
functions $\Delta_{(q)} {\cal L}|$ and $\Delta_{\phi} {\cal L}|$~\footnote{Hereafter, the vertical bar
means the restriction to the first component in the $\theta$ expansion of the corresponding superfield
expression.} and it is invariant under the component form of the hidden $N{=}4$ supersymmetry
\p{hidd88} if and only if
\be
\Delta_{(q)} {\cal L}| = -\Delta_{(\phi)} {\cal L}|  \:\:\Leftrightarrow\;\; \Delta_{(8)}{\cal L} = 0\,.
\label{CondComp}
\ee

Let us also note that the action \p{2gen} is defined up to adding to ${\cal L}$ some terms
$\omega(q, \phi)$ which simultaneously obey two dimension-4 Laplace equations
\be
\Delta_{(q)} \omega =  \Delta_{(\phi)} \omega = 0\,. \label{4harm}
\ee
Such terms do not contribute to the action, but can be used, e.g. to simplify the Lagrangian in one or
another case.

\subsection{Bosonic target metric and examples}
Using the definition \p{measure}, eqs. \p{Corol0} and the following corollaries of the latter
\be
D^{ia} \chi^A_k = -4i \delta^i_k \partial_t q^{aA}\,, \quad D^{ia}\chi^{\alpha}_b =
-4i \delta^a_b \partial_t \phi^{i\alpha}\, \lb{Corol1}
\ee
(they can be derived using the basic anticommutation relation \p{N4acomm}), it is easy to find
the purely bosonic part of the general Lagrangian \p{2gen}
\be
S^{\rm gen}_{(4+4)bos} = \frac{1}{2} \int dt \left[(\Delta_{(q)} {\cal L}|) (\partial_t q)^2 -
(\Delta_{(\phi)} {\cal L}|) (\partial_t \phi)^2\right]. \label{bos}
\ee
Note that the minus sign before the second term does not mean negative energy since the free Lagrangian
is just $\sim q^2 - \phi^2$. The $N{=}8$ supersymmetric case corresponds to the
choice
\be
S^{\rm gen}_{(8)bos} = \frac{1}{2} \int dt\, G^{(8)}\left[(\partial_t q)^2 + (\partial_t \phi)^2\right],
\quad G^{(8)}(q,\phi) = \Delta_{(q)} {\cal L}| = -\Delta_{(\phi)} {\cal L}|\,. \label{bosN8}
\ee
So the bosonic metric in this case is necessarily conformally flat (in agreement
with the ansatz proposed in \cite{GPS}) and it satisfies the dimension-8 harmonicity
condition
\be
(\Delta_{(q)} + \Delta_{(\phi)}) G^{(8)} = 0\,, \lb{ALapl}
\ee
since the Lagrangian ${\cal L}$ does. In the Subsection 2.4 the full off-shell component
structure of the superfield
Lagrangian in \p{2gen} is established. The Lagrangian \p{bos} follows from
it in the limit when all fermions
are put equal to zero.

The general $N{=}8$ supersymmetry conditions \p{Cond1} are essentially simplified for the particular
cases when the $N{=}8$ supersymmetric Lagrangian in addition possesses invariance with respect to one
or both SO(4) symmetries realized on two independent vector indices. For instance,
requiring SO(4) acting on the index $(aA)$ to be preserved restricts
$G^{i\alpha\;aA}$ to the form
\be
G^{i\alpha\;aA} = q^{aA}G^{i\alpha}(\phi, x), \quad x \equiv q^{aA}q_{aA}\,, \lb{anz1}
\ee
and ${\cal L}$ to ${\cal L}(\phi, x)$. In this particular case eqs. \p{Cond1} take the following form
\be
\mbox{(a)}\;\;\frac{\partial G^{i\alpha}}{\partial \phi^{i\beta}}
= -2\delta^\alpha_\beta\, {\cal L}_x\,, \quad
\mbox{(b)}\;\; 2G^{i\alpha} + x G^{i\alpha}
= \frac{\partial {\cal L}}{\partial \phi_{i \alpha}}\,. \lb{Cond111}
\ee
{}From the second condition follows that $G^{i\alpha}$ obeys an even stronger
condition than self-duality: it should be ``pure gauge'',
\be
G^{i\alpha} = \frac{\partial F}{\partial \phi_{i \alpha}}\,.\label{GF}
\ee
Then the set \p{Cond111} is reduced to the following two equations for the function $F$:
\be
\Delta_{(\phi)} F = -4 {\cal L}_x, \quad 2F + x F_x  = A(x) + {\cal L}\,,
\ee
$A(x)$ being an integration constant (bearing no $\phi^{i\alpha}$ -dependence).
From these equations, for any ${\cal L}$ satisfying the dimension-8 harmonicity condition \p{Harm1} one can
restore the appropriate function $F$ (up to an unessential freedom of adding solutions of
the corresponding homogeneous equation) and further  $G$ by eq. \p{GF}, thus explicitly
proving $N{=}8$ supersymmetry of the relevant $N{=}4$ superfield action.

The $N{=}8$ supersymmetry conditions are most simple in the case of full
SO(4)$\times $SO(4) invariance, in which case the Lagrangian can depend only
on the two invariants $x = q^{aA}q_{aA}$ and $y = \phi^{i\alpha}\phi_{i\alpha}$
(this case was briefly considered in \cite{ABC}). The matrix  $G^{i\alpha\;aA}$ in
this case is reduced to
\be
G^{i\alpha\;aA} = \phi^{i\alpha}q^{aA} G(x, y)
\ee
and the constraints \p{Cond1} to
\be
{\cal L}_x = G + \frac{1}{2}\,yG_y\,, \quad {\cal L}_y =
-G - \frac{1}{2}\,xG_x\,. \lb{Cond4}
\ee
The dimension-8 harmonicity equations for ${\cal L}(x,y)$ and $G^{i\alpha\;aA}$ are reduced to
\be
\mbox{(a)} \; x {\cal L}_{xx} + y {\cal L}_{yy} + 2({\cal L}_x + {\cal L}_y) = 0\,, \;\;
\mbox{(b)} \; x G_{xx} + y G_{yy} + 3(G_x + G_y) = 0\,. \lb{Lapl}
\ee

For any ${\cal L}$ satisfying (\ref{Lapl}a) it is easy to find the corresponding function
$G$ satisfying eqs. \p{Cond4} and their corollary (\ref{Lapl}b). As an example, we
explicitly present a few polynomial solutions (up to arbitrary renormalization factors)
\bea
&& {\cal L}_1 = x -y\,, \; G_1 = 1\,, \;\;{\cal L}_2 = x^2 + y^2 -3xy\,, \;
G_2 = 2(x -y)\,, \nn
&& {\cal L}_3 = x^3 - y^3 + 6xy^2 - 6x^2 y\,, \; G_3 = 3(x^2 + y^2) - 8xy\,.
\lb{Polynoms}
\eea
The first solution corresponds to the free $N{=}8$ invariant action, while the higher-order
$N{=}8$ invariants produce non-trivial sigma model bosonic metrics.

For the SO(4)$\times $SO(4) invariant case the conformal factor $G^{(8)}$ in \p{bosN8}
is related to ${\cal L}$ as
\be
\tilde{G}^{(8)} = 4\left(x{\cal L}_{xx} + 2{\cal L}_x\right) = -
4 \left(y{\cal L}_{yy} + 2{\cal L}_y \right).
\ee
The non-trivial polynomial Lagrangians in \p{Polynoms} produce the following
target bosonic metrics
\be
\tilde{G}^{(8)}_2 = 24(x -y)\,, \quad \tilde{G}^{(8)}_3 = 48(x^2 + y^2 - 3xy)\,.
\ee

It is interesting that the maximally symmetric (SO(8) invariant)
singular solution of \p{ALapl},
\be
\tilde{G}^{(8)}_{so(8)} = a_0 + a_1\frac{1}{(x + y)^3}\,,\label{so8metr}
\ee
corresponds to the following choice of the $N{=}4$ superfield Lagrangian
\be
{\cal L}_{so(8)} = \frac{a_0}{8}(x - y)
+ \frac{a_1}{16}\left(\frac{1}{x} -\frac{1}{y}\right)\frac{1}{x+y}\,. \label{Lso8}
\ee
The first term is the free Lagrangian while the second one produces a non-trivial $N{=}8, d{=}1$ sigma-model.
Note that \p{Lso8} is defined up to terms which satisfy the
dimension-4 harmonicity conditions \p{4harm}. The general function $\omega (q, \phi)$ preserving
the manifest SO(4)$\times $SO(4) symmetry is as follows
\be
\omega_{so(4)} = c_0 + \frac{c_1}{x} + \frac{c_2}{y} + \frac{c_3}{x y}\,,\label{freed}
\ee
where $c_i\,, \;i= 0, \ldots 3$ are arbitrary real constants. Using this freedom,
one can equivalently replace the second term in \p{Lso8} by the expressions
\be
\frac{a_1}{8} \frac{1}{x(x+y)} \quad \mbox{or} \quad
-\frac{a_1}{8} \frac{1}{y(x+y)}\,. \label{Equiv}
\ee

\subsection{SO(8) covariance}
Actually, the SO(8) symmetry is present already at the superfield level since it
is the hidden symmetry of the defining constraints \p{N488constr}. Two SO(4)s from this SO(8) are
manifest in the $N{=}4$ superfield description. They are formed just by two
automorphism SU(2)s realized on the indices $i, a$ and two Pauli-G\"ursey type SU(2)s
realized on the superfield indices $\alpha, A$. It turns out that the constraints
\p{N488constr} respect the covariance also under the following hidden SO(8)/[SO(4)$\times $SO(4)]
transformations
\be
\delta_{(8)} \phi^{i\alpha} = \lambda^{i\alpha\,aA} q_{aA} -
\frac{1}{2}\lambda^{l\alpha\,bA}\theta_{lb} D^{ic}q_{cA}\,, \quad
\delta_{(8)} q^{a A} = -\lambda^{k\alpha\,aA} \phi_{k\alpha} +
\frac{1}{2}\lambda^{l\alpha\,bA}\theta_{lb} D^{ia}\phi_{i\alpha}\,. \label{Hid1}
\ee
Here $\lambda^{i\alpha\,aA}$ satisfying the reality conditions with respect
to its four doublet indices encompasses just 16 parameters of the coset
SO(8)/[SO(4)$\times $SO(4)], and the additional $\theta$-dependent terms are necessary
for preserving the constraint \p{N488constr}. It is straightforward to check that the Lie bracket of
two transformations \p{Hid1} yields just the transformations of the above mentioned
manifest SU(2) symmetries, e.g.
\be
[\delta_{(8)(2)}, \delta_{(8)(1)}]\phi^{k\alpha} = \omega^{(\alpha}_{\;\;\;\beta)}\phi^{k\beta} +
\omega^{(k}_{\;\;\;l)}\phi^{l\alpha} -[
\omega^{(j}_{\;\;\;t)}\theta^{tb} +
\omega^{(b}_{\;\;\;d)}\theta^{jd}]\frac{\partial}{\partial \theta^{jb}}\phi^{k\alpha}\,,
\ee
where
$$
\omega^{(tk)} = \frac{1}{2}\left(\lambda_{(2)}^{t\gamma \,aA}\,
\lambda_{(1)}^{k}{}_{\gamma\, aA} - (1\leftrightarrow 2) \right),
$$
etc. Also, one can check that \p{Hid1} takes the manifest $N{=}4$ supersymmetry transformations into
the hidden ones \p{hidd88} and vice versa. Defining the manifest supersymmetry transformations as
\be
\delta_\epsilon \phi^{i\alpha} = -\epsilon^{kb}Q_{kb}\phi^{i\alpha}\,, \quad \delta_\epsilon q^{aA} =
-\epsilon^{kb}Q_{kb}q^{aA}\,, \quad Q_{kb} = \frac{\partial}{\partial \theta^{kb}}
- i\theta_{kb}\partial_t\,, \label{mansusy}
\ee
it is easy to find
\be
[\delta_{(8)}, \delta_\epsilon]\phi^{i\alpha} = \frac{1}{2}\eta^{\alpha A}_{(br)}D^{ia} q_{aA} =
\delta_{\eta_{(br)}}\phi^{i\alpha} \,, \quad
[\delta_{(8)}, \delta_\eta]\phi^{i\alpha} = \epsilon^{kb}_{(br)}Q_{kb}\phi^{i\alpha} =
-\delta_{\epsilon_{(br)}}\phi^{i\alpha}\,,
\ee
where
$$
\eta^{\alpha A}_{(br)} = \lambda^{l\alpha\,aA}\,\epsilon_{la}\,, \quad
\epsilon^{k b}_{(br)} = \lambda^{k\beta\,bA}\,\eta_{\beta A}\,.
$$

It is not too easy to directly check the SO(8) invariance of the full $N{=}4$ superfield action corresponding
to the Lagrangian \p{Lso8}, since its invariance is implicit (it holds modulo total spinor derivatives
under the Berezin integral), and essential use of the defining constraints \p{N488constr}
is needed when checking it.
The proof of the SO(8) invariance for the free part of \p{Lso8} is rather
simple, and it is essentially based on the identity
\be
\int dt d^4 \theta\, q^{a A}\phi^{i\alpha} = 0\,,
\ee
which follows from the constraints \p{N488constr}. The proof for the second part
of \p{Lso8} (or its equivalent forms \p{Equiv}) is rather intricate, and the more direct way
to check the SO(8) invariance of the related action is to make use of the component form
of it.

\subsection{The general component action of the multiplet (8, 8, 0)}
A direct calculation with the help of the constraints \p{N488constr} and their corollaries
\p{Corol0}, \p{Corol1}, as well as the definition of the Grassmann integration
measure \p{measure}, yields the following exact answer for the component form
of the general $N{=}4$ superfield action \p{2gen}:
\bea
S^{\rm gen}_{(8)} = \int dt {\cal L}_{comp}(q, \phi, \chi)\,, \quad
{\cal L}_{comp} = {\cal L}_{bos} + {\cal L}_{2ferm}+ {\cal L}_{4ferm}\,.  \label{Total}
\eea
Here, ${\cal L}_{bos}$ was defined in \p{bos} while two-fermion and four-fermion terms are
given by the expressions
\bea
{\cal L}_{2ferm} &=& -\frac{i}{16}\left[\nabla_t\chi_{kA}\,\chi^{kA}\,(\Delta_{(q)} {\cal L}) -
\nabla_t\chi_{a\alpha}\,\chi^{a \alpha}\,(\Delta_{(\phi)} {\cal L}) \right], \label{2ferm} \\
\nabla_t\chi_{kA} &=& \partial_t\chi_{kA} + \chi_{k}^D \,\partial_t q^c_{(D}
\,\frac{\partial \Delta_{(q)} {\cal L}}{\partial q^{c A)}}\, (\Delta_{(q)} {\cal L})^{-1}+
\chi^{i}_A\, \partial_t \phi^\alpha_{(i}
\,\frac{\partial \Delta_{(q)} {\cal L}}{\partial \phi^{k)\alpha}}\,(\Delta_{(q)} {\cal L})^{-1} \nn
&& -\,\chi^{a \alpha}\, \partial_t q_{a A}
\,\frac{\partial \Delta_{(q)} {\cal L}}{\partial \phi^{k\alpha}}\,(\Delta_{(q)} {\cal L})^{-1}
- \chi^{a \alpha} \,\partial_t \phi_{k \alpha}
\,\frac{\partial \Delta_{(\phi)} {\cal L}}{\partial q^{a A}}\,(\Delta_{(q)} {\cal L})^{-1}\,, \nn
\nabla_t\chi_{a\alpha} &=& \partial_t\chi_{a\alpha} + \chi_{a}^\beta\, \partial_t \phi^i_{(\beta}
\,\frac{\partial \Delta_{(\phi)} {\cal L}}{\partial \phi^{i \alpha)}}\,(\Delta_{(\phi)} {\cal L})^{-1} +
\chi^{d}_\alpha\, \partial_t q^A_{(d}
\,\frac{\partial \Delta_{(\phi)} {\cal L}}{\partial q^{a)A}}\,(\Delta_{(\phi)} {\cal L})^{-1} \nn
&& -\,\chi^{l B}\, \partial_t q_{a B}
\,\frac{\partial \Delta_{(q)} {\cal L}}{\partial \phi^{l\alpha}}\,(\Delta_{(\phi)} {\cal L})^{-1}
- \chi^{l B} \,\partial_t \phi_{l \alpha}
\,\frac{\partial \Delta_{(\phi)} {\cal L}}{\partial q^{a B}}\,(\Delta_{(\phi)} {\cal L})^{-1}\,, \label{2ferm2}
\eea
\bea
{\cal L}_{4ferm} &=& \frac{1}{48}\left[\frac{1}{2}\,\Omega^{(\alpha\beta)}\,\Omega_{(\alpha\beta)}
\,\Delta_{(\phi)}^2 {\cal L} - \frac{1}{2}\,\Omega^{(AB)}\,\Omega_{(AB)}\,
\Delta_{(q)}^2 {\cal L} \right. \nn
&&\left. +\, \Omega^{(\alpha\beta)}\,\chi^c_\alpha\,\chi^{iD}\,
\frac{\partial^2 \Delta_{(\phi)} {\cal L}}{\partial q^{cD} \partial \phi^{i\beta}}
- \Omega^{(AB)}\,\chi^k_B\,\chi^{\alpha b} \,
\frac{\partial^2 \Delta_{(q)} {\cal L}}{\partial q^{bA} \partial \phi^{k\alpha}} \right. \nn
&&\left.+ \, 3\,\Omega^{(\alpha\beta)}\,\Omega^{(ik)}\,
\frac{\partial^2 \Delta_{(q)} {\cal L}}{\partial \phi^{i\beta} \partial \phi^{k\alpha}}
- 3\,\Omega^{(AB)}\,\Omega^{(cd)}\,
\frac{\partial^2 \Delta_{(\phi)} {\cal L}}{\partial q^{c A} \partial q^{d B}} \right]. \label{4ferm}
\eea
Here,
\be
\Omega^{(\alpha\beta)} =\frac{1}{4}\,(\chi^\alpha_a\, \chi^{a\beta})\,, \,
\Omega^{(a b)} =\frac{1}{4}\,(\chi_\alpha^a\, \chi^{b\alpha})\,, \,
\Omega^{(AB)} =\frac{1}{4}\,(\chi^A_i\, \chi^{i B})\,, \,\Omega^{(ik)} =
\frac{1}{4}\,(\chi^i_A\, \chi^{k A})\,,
\ee
the spinors $\chi^{a\beta}$ and $\chi^{i A}$ are the lowest components of the
spinor superfields defined in \p{Corol0}, and
we omitted the vertical bar of $\Delta_{(q)} {\cal L}$ and $\Delta_{(\phi)} {\cal L}$ (now
$q^{aA}$ and $\phi^{i \alpha}$ denote the first bosonic components of the relevant
superfields).

It is rather tedious though straightforward to check that \p{Total} is invariant under the manifest $N{=}4$
supersymmetry
\be
\delta_\epsilon q^{aA} = \frac{1}{2}\epsilon^{ia}\chi^A_i\,, \; \delta_\epsilon \chi^{iA} =
-4i\,\epsilon^{ib}\partial_tq^A_b\,, \;\delta_\epsilon \phi^{i\alpha}
= \frac{1}{2}\epsilon^{ia}\chi^\alpha_a\,, \; \delta_\epsilon \chi^{b\alpha} =
-4i\,\epsilon^{ib}\partial_t\phi^\alpha_i\,,  \label{compsusy1}
\ee
without any restrictions on the functions $\Delta_{(\phi)} {\cal L}$ and $\Delta_{(q)} {\cal L}$, and
under the transformations of hidden $N{=}4$ supersymmetry
\be
\delta_\eta q^{aA} = -\frac{1}{2}\eta^{\alpha A}\chi^a_\alpha\,, \,
\delta_\eta \chi^{iA} = -4i\,\eta^{\alpha A}\partial_t\phi^i_\alpha\,, \,
\delta_\eta \phi^{i\alpha} = \frac{1}{2}\eta^{\alpha A}\chi^i_A\,, \, \delta_\eta \chi^{b\alpha} =
4i\,\eta^{\alpha A}\partial_t q^b_A\,,  \label{compsusy2}
\ee
provided only that the dimension-8 harmonicity condition holds, i.e.
\be
\Delta_{(q)} {\cal L} = -\Delta_{(\phi)} {\cal L} \equiv G^{(8)}\,.\label{8dim2}
\ee
The component transformations \p{compsusy1}, \p{compsusy2} follow from the superfield transformation
laws \p{mansusy} and \p{hidd88}. It can be directly checked that each of these transformations
closes off-shell on $t$-translations and that \p{compsusy1}, \p{compsusy2} commute with each other.

The SO(4)$\times $SO(4) invariant case corresponds to the choice ${\cal L} = {\cal L}(x, y)$ in
\p{bos}, \p{2ferm}, \p{2ferm2} and \p{4ferm}, where one should substitute the derivatives
with respect to $q^{aA}$ and $\phi^{i\alpha}$ as
\be
\frac{\partial}{\partial q^{c A}} = 2 q_{cA}\partial_x\,, \,
\frac{\partial}{\partial \phi^{i \alpha}} = 2 \phi_{i\alpha}\partial_y\,,
\Delta_{(q)} = 4\left(\partial_x + 2 x\partial_x^2 \right), \,
\Delta_{(\phi)} = 4\left(\partial_y + 2 y\partial_y^2 \right).
\ee
The superfield SO(8) transformations \p{Hid1} induce the following transformations of the component fields:
\be
\delta_{(8)} \phi^{i\alpha} = \lambda^{i\alpha\,aA}q_{aA}, \,
\delta_{(8)} q^{a B} = \lambda^{i\alpha\,aB}\phi_{i\alpha}, \,
\delta_{(8)} \chi^{a \beta} = -\lambda^{k\beta\,aA}\chi_{k A}, \,
\delta_{(8)} \chi^{i A} = \lambda^{i\beta\,aA}\chi_{a \beta}.
\ee

While the SO(8) invariance of the free version of \p{Total},
with $\Delta_{(q)} {\cal L} = -\Delta_{(\phi)} {\cal L} = const$ (it corresponds to the first term in the $N{=}8$
supersymmetric Lagrangian \p{Lso8}), is immediately seen, the check of the SO(8) invariance of the component action
corresponding to the second term in \p{Lso8} requires some effort. Nevertheless, using the properties
that in the SO(8) invariant case
\be
(\partial_x-\partial_y) G^{(8)} = 0\,, \quad {\cal L}_x +\frac{1}{4}(x + y){\cal L}_{xx} = 0, \label{Relso8}
\ee
one can check that \p{Total} is SO(8) invariant in each order in the fermionic fields. It is amusing
that for this check it suffices to use only \p{Relso8}, without resorting to the explicit form of the factor
$G^{(8)}$ (recall that $G^{(8)} \sim (x+y)^{-3}$ in the non-trivial SO(8) invariant case).

In what follows we shall argue that the ${\bf (8, 8, 0)}$ component action \p{Total} is the
generating one for the sigma-model type off-shell component actions of all other $N{=}8$ multiplets
described in \cite{ABC}:
these actions can be obtained from \p{Total} by an algorithmic procedure preserving $N{=}8$ supersymmetry.
In this sense \p{Total} is the true analog of the general sigma-model type action of the single ``root''
$N{=}4$ supermultiplet ${\bf (4, 4, 0)}$ \cite{root}. Thus the class of general actions of two
mutually ``mirror'' $N{=}4$ ``root'' multiplets   ${\bf (4, 4, 0)}$ involves a subclass of
$N{=}8$ supersymmetric  actions. The general action of this subclass given above can be called
the  ``parent'' $N{=}8$ mechanics action.

\setcounter{equation}{0}
\section{The (7,8,1) mechanics}
\subsection{Hidden $N{=}4$ supersymmetry and bosonic metric}
In the $N{=}4$ approach the multiplet ${\bf (7, 8, 1)}$
is a sum of the $N{=}4$ off-shell multiplets ${\bf (3,4,1)}$ and ${\bf (4,4,0)}$
which are described by the $N{=}4$ superfields $v^{ik} = v^{ki}$ and $q^{a A}$
subjected to the following constraints
\be
\mbox{(a)} \; D^{a(i}v^{kl)} = 0\,, \quad \mbox{(b)} \; D^{i(a} q^{b)A} = 0\,.
\lb{vqConstr}
\ee
As in the previous model, the doublet index $A$ corresponds to one external Pauli-G\"ursey type SU(2) group.
The transformations of the hidden $N{=}4$ supersymmetry completing the manifest one to
the full $N{=}8$ supersymmetry are as follows \cite{ABC},
\be
\delta_\eta v^{ik} =\frac{1}{2} \eta^{(i}_A D^{k)}_a q^{aA}\,, \quad
\delta_\eta q^{aA} = -\frac{2}{3}\eta^{iA} D^{aj}v_{ij}\,.\lb{N8vq}
\ee
The constraint (\ref{vqConstr}b) implies the same corollaries \p{Corol0}, \p{Corol1}
as in the previous case. We shall also need some consequences of the constraint
(\ref{vqConstr}a), namely
\bea
&& D^{ia} v^{jk} = \frac{1}{3}\left(\epsilon^{ji} \omega^{ak}
+ \epsilon^{ki}\omega^{aj} \right), \quad \omega^{ak} \equiv D^{ia}v^k_{\;i}\,,  \nn
&& D^{ia} \omega^{bk} = 3i \epsilon^{ab}\partial_t v^{ik}
-\frac{3i}{2}\epsilon^{ab}\epsilon^{ik} F\,, \quad F \equiv \frac{i}{6}
D_{ia}\omega^{ia}\,, \nn
&& D^{ia} F = -\frac{2}{3}\partial_t\omega^{ai}\,.\lb{Corol5}
\eea

Starting from the most general action of the superfields $v^{ik}$, $q^{aA}$ in
$N{=}4$ superspace,
\be
S^{\rm gen}_{(7)} = \int dt d^4\theta\, {\cal L}(v, q)\,, \lb{vqact}
\ee
and considering its variation under the hidden $N{=}4$ supersymmetry
transformations \p{N8vq}, one can deduce the general conditions of the
$N{=}8$ invariance of the action by requiring the variation of ${\cal L}$ to be
a total spinor derivative:
\be
\delta_\eta {\cal L} = \frac{1}{2} \eta^{(i}_A D^{k)}_a q^{aA}\frac{\partial {\cal L}}{\partial v^{ik}} -
\frac{2}{3}\eta^{iA} D^{aj}v_{ij}\frac{\partial {\cal L}}{\partial q^{aA}} =
\eta_{iA}D_{ak} G^{aA\; ik} \lb{deltaL2}
\ee
(cf. \p{deltaL1}). Equating the coefficients of the independent spinors
$\omega^{ak}$ and $\chi^{iA}$ in both sides of this condition, we find
\be
\mbox{(a)}\;\;\frac{\partial G^{aA\;ik}}{\partial q^{aB}}
= \delta^A_B \frac{\partial {\cal L}}{\partial v_{ik}}\,, \quad
\mbox{(b)}\;\;\frac{\partial G^{aA\;ki}}{\partial v^{k j}}
= -\delta^i_j \,\frac{\partial {\cal L}}{\partial q_{a A}}\,. \lb{Condvq}
\ee

As the compatibility condition of these equations, ${\cal L}$ should satisfy the
dimension-7 harmonicity equation
\be
(\Delta_{(v)} + \Delta_{(q)}){\cal L}(v,q) = 0\,, \lb{Deltavq}
\ee
where
\be
\Delta_{(v)} = \frac{\partial^2}{\partial v^{ik}\partial v_{ik}}\,.
\ee
This constraint is the only integrability condition for ${\cal L}$ following from
eqs. \p{Condvq}. From the same general $N{=}8$ supersymmetry conditions it follows that
\be
(\Delta_{(v)} + \Delta_{(q)})G^{aA\;ik} = 0\,. \lb{DeltavqG1}
\ee
For any ${\cal L}$ satisfying \p{Deltavq} the function $G^{aA\;ik}$ can be restored, up to an unessential
solution of the appropriate homogeneous equations, from \p{Condvq} and its corollary \p{DeltavqG1}.

Using the relations \p{Corol0}, \p{Corol1}, \p{Corol5} and the definition \p{measure},
one can compute the bosonic part of the most general $v, q$ action \p{vqact} as
\be
S^{\rm gen}_{(4+3)bos} = \frac{1}{2} \int dt \left\{ (\Delta_{(q)}{\cal L}|)\,(\partial_t q)^2
- (\Delta_{(v)}{\cal L}|)\left[(\partial_t v)^2 + \frac{1}{2} F^2\right]\right\}\,. \lb{bosvq}
\ee
The basic condition of $N{=}8$ supersymmetry \p{Deltavq} requires two independent
conformal factors in \p{bosvq} to coincide, implying that the general target bosonic metric in
the ${\bf (7,8,1)}$ case is conformally flat, like in the previous ${\bf (8,8,0)}$ case, i.e.
\be
S^{\rm gen}_{(7)bos} = \frac{1}{2} \int dt\, G^{(7)}\left[(\partial_t q)^2
+ (\partial_t v)^2 + \frac{1}{2} F^2\right]\,,  \lb{bosvq1}
\ee
with
\be
G^{(7)} = \Delta_{(q)}{\cal L}| = - \Delta_{(v)}{\cal L}|\,.
\ee

As we shall see later, further descending to the multiplets ${\bf (6, 8, 2)}$ and
${\bf (5,8,3)}$ preserves the fundamental properties of the conformal flatness
of the bosonic target metrics and the harmonicity of the relevant conformal factors.
It is natural to assume that these two properties are retained also upon descending
to the multiplets with a smaller number of physical bosons.
For the case of the single ${\bf (4,8,4)}$ multiplet this was independently demonstrated
in \cite{BKS} (see also \cite{BIS} where the case with few  such multiplets was considered).
In the extreme case of the multiplet ${\bf (1, 8, 7)}$ the harmonicity condition should degenerate
into its dimension-1 version which implies the corresponding conformal factor to be
a linear function of the single physical boson. This agrees with the results
of \cite{Topp1}.

To complete this Subsection, we present the particular form of the $N{=}8$
supersymmetry conditions \p{Condvq}
for the case when all three SU(2) symmetries realized on the superfields $v^{ik}$
and $q^{aA}$ are preserved. In this case ${\cal L}$ is a function of two
independent invariants $x = q^{aA}q_{aA}$ and $y = v^{ik}v_{ik}$, and the matrix
$G^{aA\;ij}$ is reduced to the form~\footnote{One could also add the term
$q^{aA} \epsilon^{ij} \tilde{G}(x, y)\,$.
However, eqs. \p{Condvq} imply for $\tilde{G}$ the homogeneous equations
$2\tilde{G} + x\tilde{G}_x =\tilde{G}_y = 0$, so this extra term is unessential and can be omitted.}
$$
G^{aA\;ij} = q^{aA} v^{ij} G(x, y)\,.
$$
Then \p{Condvq} is reduced to the simple set of two equations
\be
{\cal L}_y = G + \frac{1}{2}\,xG_x\,, \quad {\cal L}_x = -\frac{3}{4} G - \frac{1}{2} yG_y\,,
\ee
the integrability condition of which is the equation
\be
x{\cal L}_{xx} + y{\cal L}_{yy} + 2 {\cal L}_x + \frac{3}{2}\,{\cal L}_y = 0\,, \lb{Harm7}
\ee
which is just the dimension-7 Laplace equation \p{Deltavq} for the considered
particular case. It is of interest to present a few first polynomial solutions
of this equation, together with the relevant conformal factors
$\tilde{G}^{(7)} = 4(x{\cal L}_{xx} + 2 {\cal L}_x) = -4(y{\cal L}_{yy} + \frac{3}{2} {\cal L}_y)$:
\bea
&& {\cal L}_{1} = x -\frac{4}{3}y\,, \quad \tilde{G}^{(7)}_1 = 8\,, \nn
&& {\cal L}_2 = x^2 + \frac{8}{5} y^2 - 4xy\,, \quad \tilde{G}^{(7)}_2
= 24(x -\frac{4}{3}y)\,, \nn
&& {\cal L}_3 = x^3 - \frac{64}{35}y^3 - 8 x^2y + \frac{48}{5} xy^2\,, \quad
\tilde{G}^{(7)}_3 = 48(x^2 + \frac{8}{5} y^2 - 4xy)\,.\lb{Polynom2}
\eea
Thus, there exist non-trivial sigma-model-type interactions of the multiplet
${\bf (7,8,1)}\,$. The appearance of the same expressions for ${\cal  L}$ and
$\tilde{G}^{(7)}$ is due to the fact that both these quantities satisfy the dimension-7
Laplace equation and, hence, the equation \p{Harm7} in the [SU(2)]$^3$ invariant case.

\subsection{SO(7) symmetry}
An analog of the metric \p{so8metr} in the considered case is the SO(7) invariant metric
\be
\tilde{G}^{(7)}_{so(7)} = a_0 + a_1 \frac{1}{(x +y)^{5/2}}\,.
\ee
The corresponding $N{=}4$ superfield Lagrangian reads
\be
{\cal L}_{so(7)} = \frac{a_0}{8}(x -\frac{4}{3}y) + \frac{a_1}{3}\frac{1}{x\,\sqrt{x + y}}\,. \label{Lso7}
\ee
The ${\bf (7,8,1)}$ analog of the freedom \p{freed} is that of adding to \p{Lso7} the terms
\be
\tilde{\omega} = b_0 + \frac{b_1}{x} + \frac{b_2}{\sqrt{y}} +\frac{b_3}{x\,\sqrt{y}}\,.
\ee
They do not contribute to the action since they simultaneously satisfy both the dimension-4 and
dimension-3 harmonicity conditions.
In distinction to the SO(8) case, this ``gauge freedom'' is not too useful since it does not help
to bring the action \p{Lso7} to a simpler form.

The SO(7) analogs of the SO(8) superfield transformations \p{Hid1} can be also found from requiring the
SO(7) covariance of the defining  constraints \p{vqConstr},
\be
\delta_{(7)}v^{ik} = \lambda^{ik \,a A}q_{a A} -\frac{1}{2}\lambda^{l(i \,a A}\theta_{la} D^{k)c}q_{cA}\,,
\; \delta_{(7)} q^{a A} = -\lambda^{ik \,a A}v_{ik} -\frac{2}{3}\lambda^{li \,b A}\theta_{lb} D^{ta} v_{ti} \,.
\label{so7}
\ee
Here $\lambda^{ik \,a A} = \lambda^{ki \,a A}$ form 12 real parameters corresponding to the coset
SO(7)/[SO(4)$\times $SU(2)]. Like in the SO(8) case, one can easily check that eq.~\p{so7} has a correct closure
and transforms the manifest supersymmetry into the hidden one and vice versa. Its realization
on the component fields can be also easily found, but we do not explicitly quote it here.
In particular, $\delta_{(7)}F =0\,$.

\subsection{The component structure and relation to the (8, 8, 0) \break
multiplet}
The hidden $N{=}4$ supersymmetry \p{N8vq} has the following realization on the irreducible $N{=}4$ sets
of the component fields $(q^{a A}, \chi^{i A})$ and $(v^{ik}, \omega^{a i}, F)\,$, where now
we denote by the same letters the lowest components of the corresponding superfields:
\bea
&& \delta_\eta v^{ik} = -\frac{1}{2}\eta^{(i}_A\,\chi^{k)A}\,, \;
\delta_\eta q^{a A} = -\frac{2}{3}\eta^{i A}\omega^a_i\,, \; \delta_\eta \chi^{i A}
= -4i \eta^{kA}\partial_t v^{i}_{\;\;\;k} -2i \eta^{i A} F\,,  \nn
&& \delta_\eta \omega^{ak} = -3i \eta^k_A \partial_t q^{a A}\,, \qquad \delta_\eta F = \frac{1}{2}\eta^i_A
\partial_t\chi^A_i\,. \label{compHid2}
\eea

On the other hand, let us come back to the hidden supersymmetry transformations \p{compsusy2} of
the multiplet ${\bf (8, 8, 0)}$. Let us identify
there two SU(2) groups acting on the doublet indices $\alpha, \beta,...$ and $i, k,...$ and split
the field $\phi^{i\alpha}$ into its trace-free and trace part with respect to this diagonal SU(2),
\be
\phi^{i\alpha = k} = \phi^{(ik)} + \frac{1}{2}\epsilon^{ik}\phi\,. \label{Split}
\ee
Now it is easy to check that, upon identification
\be
\phi^{(ik)} = v^{ik}\,, \quad \chi^a_{\alpha = k} = \frac{4}{3}\omega^a_k\,, \quad \partial_t\phi = -F\,,
\label{Ident}
\ee
the transformation rules \p{compsusy2} precisely imply \p{compHid2}. Hence, if in the $N{=}8$ supersymmetric
component action \p{Total} (with the condition \p{8dim2}) we make the identification \p{Ident} and
assume that the conformal factor $G^{(8)}$ does not depend on $\phi$, we obtain the general
off-shell $N{=}8$ supersymmetric component action of the multiplet ${\bf (7, 8, 1)}$. The restricted
conformal factor will satisfy the dimension-7 harmonicity condition as a consequence of
the dimension-8 equation \p{8dim2}.
This consideration is the direct $N{=}8$ analog of a similar procedure in \cite{root} where the general
off-shell component action of the $N{=}4$ multiplet ${\bf (3, 4, 1)}$ was obtained from the general action of
the ``root'' multiplet ${\bf (4, 4, 0)}\,$. An essential difference is, however, that the conformal
factors in the $N{=}4$ cases generally do not satisfy any harmonicity condition, while in the
$N{=}8$ cases they  necessarily do. The automorphic duality between the multiplets
${\bf (7, 8, 1)}$ and ${\bf (8, 8, 0)}$ in the considered $N{=}4$ superfield formulation
can be depicted as
\be
{\bf (4, 4, 0)} \oplus {\bf (4, 4, 0)} \; \Rightarrow \; {\bf (4, 4, 0)} \oplus {\bf (3, 4, 1)}\,.
\label{Diag1}
\ee

In the cases of the other $N{=}8$ multiplets considered below one can also establish this ``automorphic
duality'' with the components of the ``root'' $N{=}8$ multiplet ${\bf (8, 8, 0)}$ and explicitly
obtain the full component sigma-model type actions from \p{Total}.

\subsection{Potential terms}
To close this Section, we note that for the ${\bf (7, 8, 1)}$ multiplet one can construct
the potential terms by adding to the superfield action for this multiplet a term
which is linear in the superfield $v^{(ik)}$,
\be
S^{\rm pot}_{(7)} = \frac{i}{3}m\,\int dtd^4\theta\,\eps_{ab}\, \theta^a_i \theta^b_k v^{(ik)}\,,\label{FI}
\ee
where $m$ is a real parameter of mass dimension. It is easy to see that, due to the $v^{ik}$
defining constraint (\ref{vqConstr}a), this term is invariant both
under the manifest and hidden $N{=}4$ supersymmetries and hence it is $N{=}8$ supersymmetric.
Then, integrating over $\theta$s, one finds its component ${\it off}$-${\it shell}$ form as
\be
S^{\rm pot}_{(7)} = - m\,\int dt\, F\,. \label{FI1}
\ee
Eliminating the auxiliary field $F$ in the sum of \p{FI1} and \p{bosvq1}, $F =
2m (G^{(7)})^{-1}\,$, we find the scalar on-shell potential term
\be
S^{\rm sc}_{(7)} = - m^2 \int dt\,(G^{(7)})^{-1}\,.
\ee
Thus, the single conformal factor $G^{(7)}$ specifies both the target bosonic metric
and the scalar potential. A similar mechanism of generating potential terms exists in the case
of the other $N{=}8$ multiplets which feature auxiliary fields. For the multiplet
${\bf (4, 8, 4)}$ it was demonstrated in \cite{BIS}, while for the multiplets
${\bf (6, 8, 2)}$ and ${\bf (5, 8, 3)}$ it will be shown below.~\footnote{A similar phenomenon
takes place in the case of non-linear ${\bf (4, 8, 4)}$ multiplet \cite{BKM,EI}.}

\setcounter{equation}{0}
\section{The (6, 8, 2) mechanics}
\subsection{Definitions and N=8 invariant action}
The $N{=}8$ multiplet ${\bf (6, 8, 2)}$ admits two equivalent off-shell $N{=}4$ superfield
formulations \cite{ABC}:
\be
\mbox{(a)} \;\;{\bf (6, 8, 2)} = {\bf (3, 4, 1)} \oplus {\bf (3, 4, 1)}\,, \quad
\mbox{(b)} \;\;{\bf (6, 8, 2)} = {\bf (4, 4, 0)} \oplus {\bf (2, 4, 2)}\,.  \label{26N4}
\ee
Most convenient for our purposes is the first $N{=}4$ splitting in \p{26N4}, when this multiplet
is described by two real isotriplet $N{=}4$ superfields  $v^{ik} = v^{ki}$ and  $w^{ab}= w^{ba}$ subjected
to the constraints
\bea
D^{(n}_a v^{ik)} = 0\,, \quad
D^{(c}_k w^{ab)} = 0\,. \label{33}
\eea
The superfield $v^{ik}$ is actually the same as the one used in the $N{=}4$ description of the
$N{=}8$ multiplet ${\bf (7, 8, 1)}$ in the previous Section. Like in the previous case, using \p{33}
and its corollaries given in the Appendix, one can prove that the irreducible set of the superfield projections
includes, besides
these superfields themselves, two quartets of spinors,
\be
\omega^{ia} = D^{k a} v^i_k\,, \qquad \beta^{ia} = D^{ib} w^a_b\,,
\ee
and two isosinglet auxiliary superfields,
\be
F = \frac{i}{6}\, D_{ia} \omega^{ia}\,, \quad
H = \frac{i}{6}\, D_{ia} \beta^{ia}\,.
\ee
All other superfield projections are expressed as time derivatives of these basic ones.
The hidden $N{=}4$ supersymmetry mixes the superfields $v^{(ik)}$ and $w^{(ab)}$ as
\be
\delta_\eta v^{(ik)} = -\frac{2}{3}\, \eta^{(i}_a\, D^{k)}_b w^{(ab)}\,, \qquad
\delta_\eta w^{(ab)} = \frac{2}{3}\, \eta^{(a}_i\, D^{b)}_k v^{(ik)}\,. \label{hid3}
\ee

The free action invariant under \p{hid3} reads
\be
S^{\rm free}_{(6)} = \int dtd^4\theta\,\, \Big \{v^2 - w^2 \Big \}\,. \label{Free6}
\ee

As in the previous cases, in order to find the general $N{=}8$ supersymmetric
action in the $N{=}4$ superfield formalism, one starts with the most general action of
the superfields $w^{ab}$ and $v^{ik}$,
\bea
S^{\rm gen}_{(6)} = \int dtd^4\theta\,\, {\cal L}(v,w)\,, \label{6gen}
\eea
varies the Lagrangian ${\cal L}(v,w)$ with respect to \p{hid3} and requires the variation to be
a total spinor derivative:
\be
\delta_{\eta}{\cal L}(v,w) =
-\frac{2}{3}\eta^{(i}_a\, D^{k)}_b w^{(ab)}\,\frac{\partial {\cal L}}{\partial v^{ik}} +
\frac{2}{3}\, \eta^{(a}_i\, D^{b)}_k v^{(ik)}\,\frac{\partial {\cal L}}{\partial w^{ab}}
= \eta_{i a}\, D_{k b}\, G^{ik\, ab} (v,w)\,.\label{Var}
\ee
From \p{Var}, using the defining constraints \p{33}, one finds the conditions on the function $G^{ik\, ab}$:
\be
\frac{\partial G^{ik\,a b}}{\partial v^{kn}}
= \delta^i_n\, \frac{\partial {\cal L}}{\partial w_{ab}}\,, \qquad
\frac{\partial G^{ik\, ab}}{\partial w^{bd}}
= - \delta^a_d\, \frac{\partial {\cal L}}{\partial v_{ik}}\,. \label{Constr682}
\ee
From these conditions it follows that the Lagrangian ${\cal L}$ and function $G^{ik\,a b}$
should satisfy the six-dimensional Laplace equations
\be
\Delta_{(6)} {\cal L} = 0\,, \qquad
\Delta_{(6)} G^{ik\, a b} = 0\,,
\ee
where
\be
\Delta_{(6)} = \Delta_{(v)} + \Delta_{(w)}\,,
\quad
\Delta_{(v)} = \frac{\partial^2}{\partial v^{ik} \partial v_{ik}}\,, \quad
\Delta_{(w)} = \frac{\partial^2}{\partial w^{ab} \partial w_{ab}}\,.
\ee

In the case of unbroken manifest $R$-symmetry SU(2)$\times $SU(2) the function $G^{ik\,a b}$ is reduced to
$$
G^{ik\,a b} = v^{ik}w^{ab}G (x, y)\,,
$$
where now
\be
x = v^{ik}v_{ik}\,, \; y =  w^{ab}w_{ab}\,, \label{Defxy6}
\ee
and the conditions \p{Constr682} are reduced to
\be
{\cal L}_x = -\frac{3}{4}\, G - \frac{1}{2}\,y G_y\,, \qquad
{\cal L}_y = \frac{3}{4}\, G + \frac{1}{2}\,x G_x\,.
\ee
The corresponding dimension-6 Laplace equations for the Lagrangian and
the function $G^{ik\, ab}$ in this case are reduced to the equations
\bea
&&
x {\cal L}_{xx} + y {\cal L}_{yy} + \frac{3}{2}\,\Big( {\cal L}_x + {\cal L}_y \Big) = 0\,, \label{L682} \\
&&
x G_{xx} + y G_{yy} + \frac{5}{2}\,\Big( G_x + G_y \Big) = 0\,.\label{G682}
\eea
The first few polynomial solutions of eq. \p{L682} are:
\be
{\cal L}_1 = x-y\,,\quad
{\cal L}_2 = x^2 - \frac{10}{3}\, xy + y^2\,, \quad
{\cal L}_3 = x^3 - 7 x^2y + 7 x y^2 - y^3\,. \label{Rinv}
\ee
It also bears no problem to find the relevant functions $G$ by solving eq. \p{G682}.

The general bosonic component action corresponding to the superfield action \p{6gen}
is easily found to be
\be
S^{\rm gen}_{(3+3)bos} = \frac{1}{2}\, \int dt \Big \{(\Delta_{(v)}{\cal L}|)
\left( \partial_t v^{ik} \partial_t v_{ik} + \frac{1}{2} B^2\right)-
(\Delta_{(w)}|) {\cal L}\left(\partial_t w^{ab} \partial_t w_{ab}
+ \frac{1}{2}\,H^2\right)
\Big \}\,.
\ee
The bosonic part of the $N{=}8$ supersymmetric action corresponds to the choice
\be
G^{(6)} = \Delta_{(v)}\, {\cal L} = - \Delta_{(w)} {\cal L}\,,
\ee
and so is given by
\be
S^{\rm gen}_{(6)bos} = \frac{1}{2}\, \int dt\, G^{(6)}\, \Big \{
\partial_t v^{ik} \partial_t v_{ik}
+ \partial_t w^{ab} \partial_t w_{ab}
+ \frac{1}{2}\, (B^2+H^2)
\Big \}\,, \qquad
\Delta_{(6)}\, G^{(6)} = 0\,. \label{bosN86}
\ee

In the SO(4) invariant case we have
\be
\tilde{G}^{(6)} = 4\left( x{\cal L}_{xx} + \frac{3}{2}\,{\cal L}_{x}\right) = -
4\left( y{\cal L}_{yy} + \frac{3}{2}\,{\cal L}_{y}\right). \label{so4G}
\ee
The conformal factors corresponding to the SO(4) invariant solutions \p{Rinv} are
\be
\tilde{G}^{(6)}_{1} = 4\,, \quad \tilde{G}^{(6)}_{2} = 20\left(x -y\right)\,, \quad \tilde{G}^{(6)}_{3} =
42 \left( x^2 - \frac{10}{3}\, xy + y^2 \right).
\ee

\subsection{SO(6)xSO(2) symmetry}
The constraints \p{33} reveal covariance under the two hidden mutually commuting
internal symmetry transformations
\bea
&& \delta_{(6)} v^{ik} = \lambda^{ik\,ab}\,w_{ab} -\frac{2}{3}
\lambda^{j(i\,cb}\,\theta_{j c} D^{k)a}w_{ab}\,, \nn
&& \delta_{(6)} w^{ab} = -\lambda^{ik\,ab}\,v_{ik}
+ \frac{2}{3} \lambda^{ji\,c(b}\,\theta_{j c} D^{ka)}v_{ki}\,, \label{so6}\\
&& \delta_{(2)} v^{ik} = \frac{1}{3}\,\lambda\,\theta^{(i}_b D^{k)a}w_a^b\,, \quad \delta_{(2)} w^{ab} =
-\frac{1}{3}\,\lambda\,\theta^{(a}_i D^{kb)}v_k^i\,. \label{so2}
\eea
Here, $\lambda^{ik\,ab} = \lambda^{ki\,ab} = \lambda^{ik\,ba}$ and $\lambda $ comprise, respectively,
nine and one real parameters. The transformations \p{so6} belong to the coset SO(6)/SO(4), while \p{so2}
form an abelian SO(2) group. The SO(2) transformations also commute with the manifest SO(4) symmetry.

Thus, the full $R$-symmetry in the considered case is SO(6)$\times $SO(2), as expected.
The transformations \p{so6}, \p{so2}  transform the manifest $N{=}4$ supersymmetry into the
implicit one and vice versa,
as in the previously studied cases. It is instructive to show how
these transformations are realized on the component fields $(v^{ik}, \omega^{ia}, F)$ and
$(w^{ab}, \beta^{ia}, H)$ defined as the lowest components of the corresponding superfield projections:
\bea
&& \delta_{(6)}\,v^{ik} =  \lambda^{ik\,ab}\,w_{ab}\,, \quad
\delta_{(6)}\,\beta^{ia} = \lambda^{ik\,ab}\,\omega_{kb}\,,\quad \delta_{(6)}\,F = 0\,, \nn
&& \delta_{(6)}\,w^{ab} =  -\lambda^{ik\,ab}\,v_{ik}\,, \quad
\delta_{(6)}\,\omega^{ia} = -\lambda^{ik\,ab}\,\beta_{kb}\,,\quad \delta_{(6)}\,H = 0\,, \label{so6comp}\\
&& \delta_{(2)}\,v^{ik} = \delta_{(2)}\,w^{ab} = 0\,, \quad
\delta_{(2)}\, \omega^{ia} = \frac{1}{2}\,\lambda\,\beta^{ia}\,, \;
\delta_{(2)}\, \beta^{ia} = - \frac{1}{2}\,\lambda\,\omega^{ia}\,, \nn
&& \delta_{(2)}\,F = \lambda\, H\,, \quad
\delta_{(2)}\,H = -\lambda\, F\,. \label{so2comp}
\eea
We see that the SO(6) transformations act on the physical bosonic and fermionic fields, leaving
the auxiliary fields invariant, while the SO(2) ones are realized on the fermionic and auxiliary
fields which form two SO(2) doublets. The physical bosonic fields are SO(2) singlets.

The SO(6) invariant solution of the dimension-6 harmonicity condition for the conformal
factor $G^{(6)}$ in
the bosonic action \p{bosN86} is
\be
\tilde{G}^{(6)}_{so(6)} = a_0 + a_1 \frac{1}{(x + y)^2}\,, \label{Gso6}
\ee
where now $x$ and $y$ are the lowest bosonic components of the superfields defined in \p{Defxy6}.
Using the relations \p{so4G}, it is not difficult to restore the $N{=}4$ superfield Lagrangian giving rise
to \p{Gso6} (as in the previous cases, it is defined up to adding mutual solutions of
two dimension-3 harmonicity equations).
The first term in \p{Gso6} corresponds to the free action \p{Free6}, while the second term is reproduced
from the Lagrangian
\be
{\cal L}_{so(6)} = -\frac{a_1}{2}\,\frac{1}{\sqrt{xy}}\,\arctan{\sqrt{\frac{x}{y}}}\,.\label{so6L}
\ee
Note that this Lagrangian, like the free one \p{Free6}, alters its sign under the interchange
$x \leftrightarrow y$, modulo terms $\sim const \frac{1}{\sqrt{xy}}$ which satisfy
the dimension-3 harmonicity
conditions and therefore do not contribute to the action.

\subsection{Relation to the multiplets (7, 8, 1) and (8, 8, 0)}
It is instructive to give the realization of the hidden $N{=}4$ supersymmetry \p{hid3}
on the component fields:
\bea
&& \delta_\eta\,v^{ik} = \frac{2}{3}\,\eta_a^{(i}\beta^{k)a}\,, \qquad \delta_\eta\,w^{ab}
= -\frac{2}{3}\,\eta_i^{(a}\omega^{b)i}\,, \nn
&& \delta_\eta\,\omega^{ia} = 3i\,\eta^i_b\partial_tw^{ab} - \frac{3i}{2}\,\eta^{ia} H\,, \quad
\delta_\eta\,\beta^{ia} = -3i\,\eta^a_k\partial_tv^{ik} + \frac{3i}{2}\,\eta^{ia} F\,, \nn
&& \delta_\eta\,F = \frac{2}{3}\,\eta^{ia}\partial_t\beta_{ia}\,, \qquad \delta_\eta\,H
= -\frac{2}{3}\,\eta^{ia}\partial_t\omega_{ia}\,. \label{compHidd}
\eea

Now let us compare these transformation laws with the hidden $N{=}4$ supersymmetry
transformations \p{compHid2} of
the components of the multiplet ${\bf (7, 8, 1)}\,$. The $N{=}4$ multiplet
$(v^{ik}, \omega^{ia}, F)$ is common
for both $N{=}8$ multiplets. Then let us identify the SU(2) group acting
on the doublet indices $A, B, \ldots$ of the
${\bf (4, 4, 0)}$ multiplet $(q^{aA}, \chi^{iA})$ with the SU(2) acting
on the doublet indices $a, b, \ldots$, which
implies identifying these two sets of doublet indices as well.
Then we decompose
\be
q^{b A=a} = q^{(ab)} + \frac{1}{2}\epsilon^{ba} q \label{Id2}
\ee
and identify
\be
q^{(ab)} = -w^{ab}\,, \quad  \chi^{ka} = -\frac{4}{3}\beta^{ka}\,, \quad \partial_t q = H\,.
\label{Ident2}
\ee
It is easy to check that, for the fields defined by \p{Id2} and \p{Ident2}, eqs. \p{compHid2}
imply  just the
transformation properties \p{compHidd}. In other words, we proved that the multiplets ${\bf (7, 8, 1)}$.
and ${\bf (6, 8, 2)}$ are related to each other by an automorphic duality of the same type
as the one discussed in Subsection 3.3 and relating the multiplets
${\bf (8, 8, 0)}\,$ and ${\bf (7, 8, 1)}\,$. This chain of two dualities
implies a third one directly relating the multiplets ${\bf (8, 8, 0)}\,$
and ${\bf (6, 8, 2)}\,$.
This chain can be depicted as
\be
{\bf (4, 4, 0)} \oplus {\bf (4, 4, 0)} \; \Rightarrow \; {\bf (4, 4, 0)} \oplus {\bf (3, 4, 1)}
\; \Rightarrow \; {\bf (3, 4, 1)} \oplus {\bf (3, 4, 1)}\,.  \label{Diag2}
\ee
Thus, we conclude
that the general $d{=}1$ sigma-model off-shell component action of the multiplet ${\bf (6, 8, 2)}\,$
can be obtained from the ``generating'' ${\bf (8, 8, 0)}\,$ multiplet action \p{Total}, \p{bos},
\p{2ferm}--\p{4ferm} just via (i) the proper identification of the doublet indices
$i, a, \alpha, A$; (ii) substituting the decompositions \p{Split} and \p{Id2};
(iii) assuming that the conformal
factor $G^{(8)}$ defined by \p{8dim2} bears no dependence on the fields
$\phi$ and $q$ and so can be identified with $G^{(6)}$ in \p{bosN86}; (iv) identifying
$\partial_t\phi = -F\,, \partial_t q = H\,$.

Let us make a comment on the alternative $N{=}4$ formulation (\ref{26N4}b).
It is related with the ``root''
multiplet ${\bf (8, 8, 0)}\,$ via the following chain of automorphic dualities
\be
{\bf (4, 4, 0)} \oplus {\bf (4, 4, 0)} \; \Rightarrow \; {\bf (4, 4, 0)} \oplus {\bf (3, 4, 1)}
\; \Rightarrow \; {\bf (4, 4, 0)} \oplus {\bf (2, 4, 2)}  \label{Diag22}
\ee
and can also be easily treated within our approach, giving rise to the same component Lagrangians.
Let us here present
the corresponding $N{=}4$ superfield Lagrangian yielding the SO(6) invariant metric
\p{Gso6} in components.
It now can be determined from the equations
\be
4\left( x{\cal L}_{xx} + 2\,{\cal L}_{x}\right) = -4\left( y{\cal L}_{yy} + {\cal L}_{y}\right)
= \tilde{G}^{(6)}_{so(6)}\,,
\ee
where $x= q^{aA}q_{aA}, \; y = \phi\bar\phi$, and $\phi$ is a chiral $N{=}4$ superfield describing
the multiplet ${\bf (2, 4, 2)}\,$. The Lagrangian ${{\cal L}}'{}_{so(6)}$ obtained
as a solution of these equations has the surprisingly simple form
\be
{{\cal L}}'{}_{so(6)} = \frac{a_0}{8}(x -2y) - \frac{a_1}{4}\,\frac{\log \,(x+y)}{x}\,.
\ee
By construction it satisfies the dimension-6 harmonicity condition, with the Laplacian
$$
\Delta_{(6)} = \Delta_{(q)} + 4 \frac{\partial^2}{\partial \phi \partial \bar\phi} =
4\left( x{\cal L}_{xx} + y{\cal L}_{yy} + 2\,{\cal L}_{x}  + {\cal L}_{y} \right).
$$

\subsection{Potential terms in superfields and components}

In the considered case there exist two independent Fayet-Iliopoulos-type terms preserving both
the manifest and hidden $N{=}4$ supersymmetries,
\be
S^{\rm pot}_{(6)} = \frac{i}{3}\,m \int dtd^4\theta\,\,\eps_{ab}\, \theta^a_i \theta^b_k v^{(ik)}
+ \frac{i}{3}_,\tilde m \int dtd^4\theta\,\,\eps^{ik}\, \theta^a_i \theta^b_k w_{(ab)}\,, \label{Pot6}
\ee
where $m$ and $\tilde{m}$ are two mass parameters. In components, \p{Pot6} amounts to
\be
S^{\rm pot}_{(6)} = - m \int dt\, F
 - \tilde m \int dt\, H\,.\label{pot6}
\ee
The invariance of \p{Pot6} under both $N{=}4$ supersymmetries can be checked using the
constraints \p{33}. In fact
it is simpler to check the invariance in the component formulation \p{pot6}, using
the transformation rules
\p{compHidd} (and the corresponding transformation rules of the explicit $N{=}4$ supersymmetry).
After elimination
of the auxiliary fields $F$ and $H$ from the sum of the bosonic actions \p{bosN86} and \p{pot6},
one obtains
the on-shell scalar potential term
\be
S^{\rm sc}_{(6)} = -(m^2 + \tilde{m}^2) \int dt\,\frac{1}{G^{(6)}}\,. \label{Potgen}
\ee

Note that \p{Pot6}, \p{pot6} break the SO(2) $R$-symmetry but respect the SO(6). Therefore, there exists
a unique SO(6) invariant scalar potential corresponding to the choice \p{Gso6},~\footnote{To prevent
possible confusion, let us mention that the SO(2) symmetry is broken in fermionic Yukawa-type couplings
which complete the scalar potentials to the full $N{=}8$ invariant and which are omitted
in \p{Potgen}, \p{Potso6}.}
\be
S^{\rm sc}_{so(6)} = -(m^2 + \tilde{m}^2) \int dt\,\frac{(x+y)^2}{a_1 + a_0(x + y)^2}\,.\label{Potso6}
\ee

\setcounter{equation}{0}
\section{The (5, 8, 3) mechanics}
\subsection{N=8 invariant action}

The off-shell $N{=}8$ multiplet ${\bf(5, 8, 3)}$ admits two equivalent $N{=}4$
superfield splittings \cite{ABC},
\be
\mbox{(a)}\;\;{\bf(5, 8, 3)} = {\bf(4, 4, 0)} \oplus {\bf(1, 4, 3)}\,,
\quad \mbox{(b)}\;\;{\bf(5, 8, 3)} = {\bf(3, 4, 1)} \oplus  {\bf(2, 4, 2)}. \label{TwoN45}
\ee
For reasons to be explained below we choose the option $\mbox{(a)}$ and will describe
the multiplet in question
by the $N{=}4$ superfields $q^{aA}$ and $u$ subjected to the constraints
\bea
D^{k (b} q^{a) A} = 0\,, \qquad
D^{k (a} D^{b)}_k u = 0\,. \label{constrqu}
\eea
The superfield $q^{aA}$ is the same as in Sections 2 and 3, and we will use the same
notation for its irreducible
superfield projections. Thus, besides the superfields $q^{aA}$ and $u$ themselves, the constraints
\p{constrqu} single out the following independent projections, the lowest components
of which constitute the irreducible field content of the multiplet ${\bf(5, 8, 3)}\,$.
\vspace{0.3cm}

\noindent{\it Spinors}:
\be
\xi^{k b} = D^{k b} u \,, \qquad \chi^{k A} = D^{k b} q^{A}_b\,.\label{Spin}
\ee
\vspace{0.3cm}

\noindent{\it Auxiliary fields}:
\be
A^{(ik)} = - \frac{i}{2}\, D^{(k}_b \xi^{i) b}\,.\label{Aux}
\ee

To prove that any other superfield projection is expressed as a time derivative of these basic ones,
one can
use the identities \p{Corol0}, \p{Corol1} together with those for the superfield $u$,
\bea
&&
D^{i a} D^{k b} u = D^{i a} \xi^{k b}
= i \eps^{ik} \eps^{ab} \partial_t u + i \eps^{ab} A^{(ik)}\,, \label{Ident-u}\\
&&
D^{l a} A^{(ik)} = i \eps^{kl} \partial_t \xi^{ia}
+ i \eps^{il} \partial_t \xi^{ka}\,.\label{u-identit}
\eea

The hidden $N{=}4$ supersymmetry is realized by the transformations
\be
\delta_\eta q^{a A} = \eta_{k}^A\, D^{ka} u\,, \qquad
\delta_\eta u = \frac{1}{2}\, \eta_{k A}\, D^{k a} q_a^{A}\,.\label{Hidd5}
\ee

The invariant free action is given by
\be
S^{\rm free}_{(5)} = \int dtd^4\theta\,\, \Big \{q^2 - 2\, u^2 \Big \}\,.\label{Free5}
\ee

The generic action of the two $N{=}4$ superfields considered,
\bea
S^{\rm gen}_{(5)} = \int dtd^4\theta\,\, {\cal L}(q,u)\,, \label{Genqu}
\eea
possesses $N{=}8$ supersymmetry if, under the transformations \p{Hidd5}, it transforms into a total
spinor derivative,
\be
\delta_\eta {\cal L}(q,u) = \eta_{k}^A\,D^{ka} u \frac{\partial {\cal L}}{\partial q^{aA}} +
\frac{1}{2}\,\eta_{k A}\, D^{k a} q_a^{A}\frac{\partial {\cal L}}{\partial u} =
\eta_{i A}\, D_{k b}\, G^{ik\, bA} (q,u)\,.
\ee
By comparing the coefficients in the left- and right-hand sides of this equality, one finds the general
conditions of $N{=}8$ supersymmetry in the system under consideration, namely
\be
\frac{\partial G^{ik\, a A}}{\partial u}
= \eps^{ik}\,\frac{\partial {\cal L}}{\partial q_{a A}}\,, \qquad
\frac{\partial G^{ik\, a A}}{\partial q^{a B}}
= - \eps^{ik}\, \delta^{A}_{B}\, \frac{\partial {\cal L}}{\partial u}\,.\label{N8cond5}
\ee
These conditions can be essentially simplified because they imply that,
up to an unessential constant,
\be
G^{ik\, a A} = \eps^{ik}\, G^{a A}\,.
\ee
Then, \p{N8cond5} are equivalently rewritten as
\be
\frac{\partial G^{aA}}{\partial u} = \frac{\partial {\cal L}}{\partial q_{a A}}\,, \quad
\frac{\partial G^{a A}}{\partial q^{a B}}
= - \delta^{A}_{B}\, \frac{\partial {\cal L}}{\partial u}\,.\label{N8cond52}
\ee

As in the previous cases, the only constraint on the Lagrangian ${\cal L}$, following from
the set of equations \p{N8cond52} as their compatibility condition, is the dimension-5 harmonicity
equation
\be
\Delta_{(5)}\, {\cal L} = 0\,,
\ee
where now
\be
\Delta_{(5)} = \Delta_{(q)} + \Delta_{(u)}\,,
\quad
\Delta_{(q)} = \frac{\partial^2}{\partial q^{a A} \partial q_{a A}}\,, \quad
\Delta_{(u)} = 2\,\frac{\partial^2}{\partial u^2}\,.
\ee
Eqs. \p{N8cond52} also imply the same equation for the function $G^{aA}$, i.e.
\be
\Delta_{(5)} G^{aA} = 0\,.
\ee

As in the previous cases, all these equations are drastically simplified for
$R$-symmetric Lagrangians. If we wish to preserve all three involved SU(2) symmetries,
we are led to choose the ansatz
$$
G^{aA} = q^{aA} G(x, u)\,, \quad x = q^{aA}q_{aA}\,,
$$
for which eqs. \p{N8cond52} are reduced to
\be
G_u = 2 {\cal L}_x \,, \qquad
2G +  x G_x = -{\cal L}_u \,. \label{Ansatz}
\ee
The corresponding equations for ${\cal L}$ and $G$, derived as the integrability conditions of the
system \p{Ansatz}, have the following form:
\bea
x G_{xx} + 3\, G_x +\frac{1}{2}\, G_{u u} = 0\,, \qquad
x {\cal L}_{xx} + 2 {\cal L}_x +\frac{1}{2}\, {\cal L}_{uu} = 0\,. \label{so4inv}
\eea
The second equation is just the dimension-5 Laplace equation for ${\cal L}$ in the variables $x, u\,$.
The first few polynomial solutions of this linear equation read
\be
{\cal L}_1 = x-2 y\,, \quad
{\cal L}_2 = x^2 - 6 xy +2y^2\,, \quad
{\cal L}_3 = x^3 - 12 x^2y + 12 x y^2 - \frac{8}{5} y^3\,,
\ee
where we defined, by analogy with the previously studied cases, $y = u^2\,$.
The first solution is the free action, the remaining ones generate non-trivial sigma-model-type
interactions.

The bosonic component action obtained from the general $N{=}8$ supersymmetric action
of the superfields $q^{aA}, u$ reads
\be
S^{\rm gen}_{(5)bos} = \frac{1}{2}\, \int dt\, G^{(5)}\, \Big \{
\partial_t q^{a B} \partial_t q_{a B}
+ \partial_t u \partial_t u
+ \frac{1}{2}\, A^{(ik)} A_{(ik)}
\Big \}\,, \qquad
\Delta_{(5)}\, G^{(5)} = 0\,,\label{bos5}
\ee
where
\be
G^{(5)} = \Delta_{(q)}\, {\cal L} = - \Delta_{(u)} {\cal L}\,.
\ee
For the SO(4) invariant case \p{so4inv} the conformal factor $\tilde{G}^{(5)}$ is related to ${\cal L}(x, y)$
by
\be
\tilde{G}^{(5)} = 4(x {\cal L}_{xx} + 2 {\cal L}_x) = -4 ({\cal L}_{y} + 2 y{\cal L}_{yy})\,.\label{Gso4}
\ee

\subsection{SO(5)xSU(2) symmetry}
The $N{=}4$ superfield constraints \p{constrqu} are preserved under the following hidden internal
symmetry transformations \cite{BIKL2},
\be
\delta_{(5)} q^{a A} = \lambda^{aA}\, u - \lambda^{c A}\,\theta_{kc} D^{ka}u\,, \quad
\delta_{(5)} u = -2\lambda^{aA}\, q_{aA} + \frac{1}{2} \lambda^{c A}\,\theta_{kc} D^{kb}q_{bA}\,.
\label{Hidd5Int}
\ee
Here, $\lambda^{aA}$ is a quartet of real parameters. It is straightforward to check that these
transformations
close on SO(4) $=$ SU(2)$_R\times $SU(2) where the first SU(2) factor acts on the doublet
indices $a$ of the superfield $q^{aA}$ and the Grassmann coordinate $\theta^{k a}$, i.e. it belongs
to the manifest
$R$-symmetry group SO(4)$_{R}\,$, while the second SU(2) factor acts on the extra doublet
index $A$ of $q^{a A}$ and so
commutes with the manifest $N{=}4$ supersymmetry. Thus the transformations \p{Hidd5Int}, together
with the two SU(2) groups
just mentioned, form the 10-parameter group SO(5), and these transformations  belong to
the coset SO(5)/SO(4).
It can be also checked that the second manifest $R$-symmetry group,  SU(2)$_{R'}$ acting on the doublet
indices $i, k$, commutes with \p{Hidd5Int}.  Thus the full $R$-symmetry of the considered case
is SO(5)$\times $SU(2)$_{R'}\,$. It is instructive
to give the realization of \p{Hidd5Int} on the component fields, using the definitions
\p{Spin}, \p{Aux}, the relation
\p{Ident-u} and the constraint \p{N488constr} with its corollary \p{Corol1}:
\bea
&& \delta_{(5)} q^{a A} = \lambda^{aA}\, u\,, \qquad \delta_{(5)} u = -2\lambda^{aA}\, q_{aA}\,,
\label{compSO51} \\
&& \delta_{(5)} \chi^{iA}=- 2\lambda^{b A}\xi^i_b\,, \quad \delta_{(5)}\xi^{ia} =
-\frac{1}{2} \lambda^{a A}\chi^{i}_A\,, \quad \delta_{(5)} A^{ik} = 0\,. \label{compSO52}
\eea
We observe, in particular, that the auxiliary field $A^{ik}$ is inert under the SO(5) group and
is transformed only under the automorphism SU(2) group which acts on the indices $i, k, \ldots\,$
and commutes
with SO(5). The physical bosons are transformed only under SO(5). On the fermions, like
in the previous cases,
the whole $R$-symmetry group SO(5)$\times $SU(2) is effective.

As in the previous models, it is interesting to establish the explicit form of
the $N{=}4$ superfield
Lagrangians corresponding to the bosonic metric with maximal internal symmetry. In the case
under consideration,
this is SO(5) symmetry, and the SO(5) invariant solution of the dimension-5 harmonicity equation for
the conformal factor $G^{(5)}$ in \p{bos5} reads
\be
\tilde{G}^{(5)}_{so(5)} = a_0 + a_1 \frac{1}{(2x + y)^{3/2}}\,, \qquad x = q^2, \;\; y = u^2\,,
\label{G5so5}
\ee
where, as in the previous cases, $a_0, a_1$ are some real constants. From the transformation
rule \p{compSO51}
it follows that $ 2x + y = 2q^{aA}q_{aA} + u^2$ is indeed SO(5) invariant. Using the
definition \p{Gso4}, one finds
that the first term in \p{G5so5} corresponds to the free Lagrangian \p{Free5},
while the second term is reproduced
from the Lagrangian
\be
{\cal L}_{so(5)} = -\frac{a_1}{4}\,\frac{\sqrt{2x + y}}{x}\;. \label{LagrSO5}
\ee
As in the previous cases, this Lagrangian is defined modulo
\be
c_0 + c_1\, u + c_2\,\frac{1}{x} + c_3\frac{u}{x}\,,
\ee
which simultaneously solves the dimension-3 and dimension-1 Laplace equations and so
does not contribute into the action. Choosing, e.g.,
\be
c_0 = c_1 = c_2 = 0\,, \quad c_3 = \frac{a_1}{4}\,,
\ee
one can cast the Lagrangian \p{LagrSO5} into the alternative form
\be
{\cal L}_{so(5)} = \frac{a_1}{4}\,\frac{u - \sqrt{2x + y}}{x}
= -\frac{a_1}{2}\,\frac{1}{u + \sqrt{2x + y}}\,. \label{LagrSO51}
\ee
It coincides with the $N{=}8$ superconformally invariant  Lagrangian
found in \cite{BIKL2} by a rather complicated recurrence procedure, starting from some general $N{=}4$
superconformally invariant action for the superfields $q$ and $u$ and then
requiring invariance under the hidden $N{=}4$ supersymmetry. We see that the same action
can be recovered
much more simply from the requirement of SO(5) invariance. Also, it follows from our consideration
that the $N{=}8$ supersymmetric Lagrangians \p{LagrSO5} and \p{LagrSO51} automatically satisfy
the dimension-5
harmonicity condition, a property which was not noticed in \cite{BIKL2}. We remark that this paper
also constructed an $N{=}8$ superconformal Lagrangian for the alternative $N{=}4$
splitting of the multiplet
${\bf (5, 8, 3)}$ corresponding to the option $\mbox{(b)}$ in \p{TwoN45}. It can be written as
\be
{\cal L}_{(b)}^{conf} \;\sim \; \frac{\log\left(\sqrt{W^2}
+ \sqrt{W^2 + \Phi\bar\Phi}\right)}{\sqrt{W^2}}\,,
\label{SUpc}
\ee
where $W^2 = W^{ab} W_{ab}$ and $W^{ab}$ represents the multiplet ${\bf (3, 4, 1)}$, while
$\Phi$ and $\bar \Phi$ are chiral and antichiral $N{=}4$ superfields describing
the multiplet ${\bf (2, 4, 2)}$.
We have checked that this Lagrangian can also be recovered from the requirement of SO(5) invariance,
and that it satisfies the appropriate dimension-5 harmonicity condition
$$
\left(\frac{\partial^2}{\partial W^{ab}\partial W_{ab}}
+ 4 \frac{\partial^2}{\partial \Phi\partial\bar\Phi }
\right){\cal L}_{(b)}^{conf} = 0\,.
$$
This confirms the universal character of the harmonicity conditions for the $N{=}4$ Lagrangians as
the conditions of $N{=}8$ supersymmetry.

\subsection{Relation to the multiplet (8, 8, 0)}
Let us now demonstrate that the $N{=}8$ multiplet ${\bf (5, 8, 3)}$ in the $N{=}4$ superfield
description (\ref{TwoN45}a) is,
by a variant of the automorphic duality, directly related to the ``root'' $N{=}8$ multiplet
${\bf (8, 8, 0)}\,$. This implies that its general off-shell sigma-model component action can be
restored by simple rules from the ``parent''  ${\bf (8, 8, 0)}$ action,
as in the cases considered in the previous Sections.

To this end, let us first write the realization of the hidden $N{=}4$ supersymmetry \p{Hidd5} on the component
fields:
\bea
&&\delta_\eta q^{aA} = \eta^A_k\,\xi^{ka}\,, \qquad  \delta_\eta u = \frac{1}{2}\,
\eta_{kA}\,\chi^{kA}\,, \nn
&&\delta_\eta \chi^{kA} = 2i\,\left(\eta^{kA}\partial_t u + \eta^{A}_l A^{lk}\right), \;
\delta_\eta \xi^{ka} = -2i\,\eta^{k}_{A}\partial_t q^{aA}\,, \;  \delta_\eta A^{ik} =
-\eta^{(i}_{A}\,\partial_t\chi^{k)A}\,. \label{Hidd5comp}
\eea
As the second step, let us identify, in the hidden supersymmetry transformations  of the component
fields of the
${\bf (8, 8, 0)}$ multiplet \p{compsusy2}, two SU(2) groups realized on the indices $i$ and $\alpha$
of the second
${\bf (4, 4, 0)}$ multiplet, and decompose the field $\phi^{i\,\alpha = k}$ just on the pattern of \p{Split},
\be
\phi^{ik} = \phi^{(ik)} +\frac{1}{2}\,\epsilon^{ik} \,\phi\,, \quad \phi = \epsilon_{ki}\phi^{ik}\,.
\ee
Now it is easy to check that after the formal identifications
\be
\phi = u\,, \quad \chi^{\alpha=k\,a} = 2\xi^{k a}\,, \quad \partial_t \phi^{(ik)} = \frac{1}{2}\,A^{ik}
\label{Autom85}
\ee
the transformations \p{compsusy2} precisely yield \p{Hidd5comp}.

It immediately follows from this remarkable fact that the general sigma-model type off-shell action
of the multiplet ${\bf (5, 8, 3)}$ can be obtained from the ${\bf (8, 8, 0)}$ multiplet
action \p{Total}, \p{bos}, \p{2ferm}--\p{4ferm} just by making the substitutions \p{Autom85} in the latter and
by eliminating the dependence on the triplet $\phi^{(ik)}$ in the general conformal factor $G^{(8)}$
defined by \p{8dim2}. The reduced factor satisfies the dimension-5 harmonicity condition, as expected,
and hence it can be identified with $G^{(5)}$ (recall \p{bos5}).

The direct automorphic duality between the off-shell $N{=}8$ multiplets ${\bf (8, 8, 0)}$
and ${\bf (5, 8, 3)}$  can be
depicted as
\be
{\bf (4, 4, 0)} \oplus {\bf (4, 4, 0)} \,\Rightarrow \, {\bf (4, 4, 0)} \oplus {\bf (1, 4, 3)}\,.
\label{Diag3}
\ee
In contrast to the chain of dualities \p{Diag2}, this chain does not go through any intermediate step.
Nevertheless, there exists another option, also finally resulting in the multiplet ${\bf (5, 8, 3)}$,
but in the alternative $N{=}4$ superfield description (\ref{TwoN45}b)
including the chiral $N{=}4$ supermultiplet:
\bea
&& {\bf (4, 4, 0)} \oplus {\bf (4, 4, 0)} \; \Rightarrow \; {\bf (4, 4, 0)} \oplus {\bf (3, 4, 1)}
\; \Rightarrow \; {\bf (3, 4, 1)} \oplus {\bf (3, 4, 1)}  \nn
&& \Rightarrow \;
{\bf (3, 4, 1)} \oplus {\bf (2, 4, 2)}\,. \label{Diag4}
\eea
This is just a continuation of the chain \p{Diag2}. The final component action is of course equivalent to
the one produced by the chain \p{Diag3}.

\subsection{Potential terms}
For the case under consideration,
the ${\it off}$-${\it shell}$ $N{=}8$ supersymmetric potential terms in superfield and component form read
\bea
&& S^{\rm pot}_{(5)} = m\,\int dtd^4\theta\,\eps_{ab}\, \theta^a_i \theta^b_k C^{(ik)} u\,,
\label{Pot5} \\
&& S^{\rm pot}_{(5)comp} = -m\,\int dt\,C_ {(ik)} A^{(ik)}\,. \label{pot5}
\eea
They clearly break the SU(2) factor in the full $R$-symmetry group SO(5)$\times $SU(2)
down to U(1) while preserving the SO(5) factor. After eliminating $A^{ik}$ from the full action
by its algebraic equations of motion there arises an on-shell scalar potential term accompanied by some Yukawa-type
fermionic couplings. The scalar potential can be found just from the sum of \p{bos5} and \p{pot5},
\be
S ^{\rm sc}_{(5)} = -m^2 C^2 \int dt\, (G^{(5)})^{-1}\,.
\ee
Though this term does not exhibit any breaking of the $R$-symmetry SU(2) (because only
the square $C^2= C^{ik}C_{ik}$
appears), the suppressed fermionic terms explicitly involve $C^{ik}\,$, and so
this SU(2) is broken in the total on-shell
action.

Finally, we would like to point out that the general $N{=}4$ superfield action for
the $N{=}8$ multiplet ${\bf (5, 8, 3)}$
in the splitting ${\bf (3, 4, 1)} \oplus {\bf (2, 4, 2)}$ was constructed previously and studied
in \cite{DE,AnSm,ISm}.
At the same time, the alternative splitting ${\bf (4, 4, 0)} \oplus {\bf (3, 4, 1)}$ was not
elaborated on too much.
In this paper, we have filled this gap and have shown that it yields equivalent results for the action
and potentials. We also
established the precise relation of the multiplet ${\bf (5, 8, 3)}$
to the ``root'' $N{=}8$ multiplet ${\bf (8, 8, 0)}$ through an automorphic duality,
which was not explicitly done before.

\setcounter{equation}{0}
\section{Summary, further examples and outlook}
In this paper we have studied several so far unexplored models of $N{=}8$ supersymmetric mechanics with
${\bf 8}$ physical fermions \cite{ABC} in the off-shell $N{=}4$ superfield approach: those associated
with the $N{=}8$ multiplets ${\bf (8, 8, 0)}$, ${\bf (7, 8, 1)}$ and ${\bf (6, 8, 2)}\,$. Also,
the model associated with the multiplet ${\bf (5, 8, 3)}$ was studied for the $N{=}4$ splitting
${\bf (5, 8, 3)} = {\bf (4, 4, 0)} \oplus {\bf (1, 4, 3)}$ which was not addressed in full generality
before. We derived general conditions of the existence of the second hidden $N{=}4$ supersymmetry
in these models, gave the corresponding general superfield actions and their component bosonic cores
(and the full component action for the ${\bf (8, 8, 0)}$ case), considered some instructive
examples including the actions with the maximal internal symmetry
SO$({\bf b})\times$ SO$({\bf a})$ where ${\bf b}$ and ${\bf a}$ are the
corresponding numbers of the physical and auxiliary bosonic fields, and
found the precise realization of this maximal symmetry on the superfields
and component fields. Three main characteristic common features of the models studied
here can be summarized as follows.

\begin{itemize}
\item
The general condition ensuring $N{=}8$ supersymmetry of the corresponding $N{=}4$ superfield
Lagrangians is that the latter should satisfy the harmonicity condition with respect to
the involved bosonic $N{=}4$ superfields as arguments.
\item
The bosonic target metric is always conformally flat, with the conformal factor being related
in a simple way to the
superfield Lagrangian and obeying the same dimension-${\bf b}$ harmonicity condition. For the multiplets
with auxiliary
fields, the same conformal factor specifies the on-shell scalar potential of physical
bosonic fields.
\item
The $N{=}8$ mechanics models reveal an interesting hierarchical structure. All these are related
to the ``root''
${\bf (8, 8, 0)}$ model through the ``automorphic duality'' which is the opportunity to replace,
without affecting
$N{=}8$ supersymmetry, time derivatives of some physical bosonic fields in a given off-shell
multiplet by the auxiliary field with the same transformation rule, thus producing a new
off-shell multiplet. Using this duality, the off-shell
component sigma-model type actions of all considered multiplets can be
recovered by simple rules from the general component action of the multiplet ${\bf (8, 8, 0)}$
which is thus the generating action for all models.
\end{itemize}

All these properties seem to extend also to the $N{=}8$ mechanics models associated
with the rest of the multiplets
classified in \cite{ABC}, viz. the multiplets ${\bf (4, 8, 4)}$, ${\bf (3, 8, 5)}$, ${\bf (2, 8, 7)}$,
${\bf (1, 8, 7)}$ and ${\bf (0, 8, 8)}\,$. Actually, our analysis can be directly applied
to any $N{=}4$ splitting of these multiplets, such that the involved $N{=}4$ superfields
possess manifestly $N{=}4$ supersymmetric actions in the ordinary
$N{=}4, d{=}1$ superspace. The latter criterion fails to be valid for the purely fermionic
$N{=}4$ multiplet
${\bf (0, 4, 4)}$ the actions of which can be formulated in the ordinary $N{=}4$ superspace
only with the inclusion of
explicit Grassmann coordinates. Therefore, the $N{=}4$ splittings involving this multiplet, e.g.
${\bf (4, 8, 4)} = {\bf (4, 4, 0)} \oplus {\bf (0, 4, 4)}$, fall beyond our analysis.
Since the $N{=}4$ splittings
of the multiplets ${\bf (1, 8, 7)}$ and ${\bf (0, 8, 8)}\,$ are unique and necessarily
include the multiplet
${\bf (0, 4, 4)}$ \cite{ABC}, our analysis cannot be directly applied to these special cases.
Nevertheless,
it is reasonable to assume that at least two last characteristic features from the above list
are also shared by the models
related to these two multiplets. Indeed, it was argued in ref. \cite{Topp1} in a different
(component) approach  that
the bosonic target metric in the ${\bf (1, 8, 7)}$ case should be at most linear in
the physical bosonic field, which is just the general solution of the dimension-1
``Laplace equation'' $\Delta_{(1)}G^{(1)} = 0\,$. As for the ``extreme'' fermionic
multiplet ${\bf (0, 8, 8)}\,$, its general action should coincide with the free one;
however, containing $8$ auxiliary fields, this multiplet could produce new
non-trivial scalar potentials while being coupled to $N{=}8$ multiplets
featuring physical bosonic fields.

Concerning the multiplet ${\bf (4, 8, 4)}$ in the splitting ${\bf (4, 4, 0)} \oplus {\bf (0, 4, 4)}$,
it is worth noting that in ref. \cite{BKS} it was considered in the harmonic $N{=}4, d{=}1$
superspace \cite{N4HSS}
where both its $N{=}4$ constituents (including their off-shell actions) admit a manifestly
$N{=}4$ supersymmetric description.~\footnote{Actually, {\it all} $N{=}4$ off-shell multiplets
admit natural
formulations in the $N{=}4, d{=}1$ harmonic superspace \cite{N4HSS,DI1,DI2} which so seems
to supply the most adequate arena
for treating these multiplets.} The general component action of the relevant $N{=}8$ mechanics
model was found
to show the second property from the list above.

In fact, this $N{=}8$ multiplet admits the alternative $N{=}4$ splittings
\be
\mbox{(a)} \;\;{\bf (4, 8, 4)}= {\bf (3, 4, 1)} \oplus {\bf (1, 4, 3)} \quad \mbox{and} \quad
\mbox{(b)} \;\; {\bf (4, 8, 4)}= {\bf (2, 4, 2)} \oplus {\bf (2, 4, 2)}\,, \label{484}
\ee
which nicely match with our approach and naturally continue the automorphic duality
chains \p{Diag3} and \p{Diag4}.
Though we did not fully explore the $N{=}8$ systems associated with these options, it
is obvious that they are well inscribed into the hierarchical structure of the $N{=}8$ mechanics
models. Therefore, they have to exhibit all
the basic features listed above and to give rise to actions equivalent to the ones constructed
in \cite{BKS} (see also \cite{BIS}).

As an example of how our approach works in this case, let us present the $N{=}8$ supersymmetric Lagrangian
of the ${\bf (3, 4, 1)}$ and ${\bf (1, 4, 3)}$ superfields $v^{ik} = v^{ki}$ and $u$ which yields the maximally
symmetric target metric satisfying the dimension-4 harmonicity condition, this time the SO(4) invariant one,
\be
\tilde{G}^{(4)}_{so(4)} = a_0 + a_1 \frac{1}{x + y}\,, \qquad x = v^{ik}v_{ik}\,, \;\; y = u^2\,.\label{484a}
\ee
Here, the term $a_0$ corresponds to the free Lagrangian, while the term $\sim a_1$ belongs to a sigma-model
Lagrangian with a non-trivial SO(4) invariant self-interaction. The appropriate
Lagrangian ${\cal L}_{so(4)}$
is defined by the equation
\be
\tilde{G}^{(4)}_{so(4)} = 4\left(x{\cal L}_{xx} + \frac{3}{2}{\cal L}_{x}\right) =
-2\left({\cal L}_{y} + 2y{\cal L}_{yy}\right)
\ee
and, up to the freedom of adding solutions of the dimension-3 and dimension-1 Laplace equations,
is given by
the expression
\be
{\cal L}_{so(4)} = \frac{a_0}{6}\left(x - 3y\right)
- a_1\,\left[\sqrt{\frac{y}{x}}\,\arctan \sqrt{\frac{y}{x}} - \frac{1}{2}\,\log\,(x+y)\right].  \label{484b}
\ee
Note that the $N{=}8$ bi-harmonic superspace action of the multiplet ${\bf (4, 8, 4)}$ corresponding to
the SO(4) invariant metric \p{484a} was found in ref. \cite{BIS}.

Analogously, for the multiplet  ${\bf (3, 8, 5)}$ there also exists an $N{=}4$ superfield splitting
which allows treatment by our method, namely
\be
{\bf (3, 8, 5)}= {\bf (2, 4, 2)} \oplus {\bf (1, 4, 3)}\,. \label{385}
\ee
It continues the automorphic duality chain \p{Diag4} and \p{484}. For the corresponding metric with
the maximal symmetry of the SO(3),
\be
\tilde{G}^{(3)}_{so(3)} = b_0 + b_1 \frac{1}{\sqrt{x + y}}\,, \qquad x = \phi\bar\phi\,, \; y = u^2\,,
\ee
where $\phi$ is the lowest component of the $N{=}4$ chiral superfield representing the
multiplet ${\bf (2, 4, 2)}$,
one can also find the relevant $N{=}4$ superfield Lagrangian from the relation
\be
\tilde{G}^{(3)}_{so(3)}  = 4\left(x{\cal L}_{xx} +{\cal L}_{x}\right)
= -2\left({\cal L}_{y} + 2y{\cal L}_{yy}\right).
\ee
The appropriate solution is
\be
{\cal L}_{so(3)} = \frac{b_0}{4}\left(x - 2y\right) + b_1\left[ \sqrt{x + y} -
\sqrt{y}\,\log \left(\sqrt{y} +\sqrt{x+y}\right) \right]. \label{385a}
\ee
The piece $\sim b_1$ in \p{385a} is nothing but the $N{=}8$ superconformal action of the multiplet
${\bf (3, 8, 5)}$ found in \cite{BIKL2} by a rather involved recurrence procedure. We see how simple
is its derivation in our framework.

A final example supporting the general character of the hierarchical structure of
the $N{=}8$ mechanics model
is the multiplet ${\bf (2, 8, 6)}$, which has a unique $N{=}4$ splitting appropriate
for effecting our procedure,
\be
{\bf (2, 8, 6)}= {\bf (1, 4, 3)} \oplus {\bf (1, 4, 3)}\,. \label{385b}
\ee
Clearly, this splitting is the next link in the automorphic duality chain \p{Diag4} continued to \p{385}.
The corresponding maximally symmetric metric solving the dimension-2 harmonicity condition
is the SO(2) invariant one
\be
\tilde{G}^{(2)}_{so(2)} = c_0 + c_1\,\log (u^2 + v^2)\,,
\ee
where $u$ and $v$ are the first components of the two ${\bf (1, 4, 3)}$ superfields (which are ``mirror''
to each other \cite{ABC}). The corresponding $N{=}4$ Lagrangian ${\cal L}_{so(2)}$ is determined
from the set of equations
\be
{\cal L}_{uu} = - {\cal L}_{vv} = \tilde{G}^{(2)}_{so(2)}
\ee
and, modulo harmless solutions of two dimension-1 ``Laplace equations'', is given by the expression
\be
{\cal L}_{so(2)} = \frac{(c_0 -3c_1)}{2}\,(u^2 - v^2) + \frac{c_1}{2}\left[(u^2 -v^2)\,\log (u^2 +v^2) +
4vu\,\arctan \,\frac{u}{v}\right].
\ee
Note that both terms in this expression are odd with respect to the permutation $u \leftrightarrow v$
(up to terms vanishing under the $N{=}4$ superspace integral).

The results of this paper raise some questions and suggest several further directions of study.

First of all, it would be interesting to work out in more detail some of the models studied here, both at
the classical and quantum level, for specific choices of the metric factor, and to establish their links
with systems of current interest, e.g. superextensions of integrable Calogero-Moser models.
In connection with the AdS$_2$/CFT$_1$ version of the general ``string/gauge'' correspondence and
the supersymmetric black-hole story it is also of primary interest to study superconformally invariant
$N{=}8, d{=}1$ models. Clearly, the non-trivial
$N{=}4$ superfield Lagrangians with maximal internal symmetry constructed here for many multiplets
and $N{=}4$ splittings
are just candidates for the appropriate superconformal mechanics actions. This is supported by the
fact that the
nonlinear Lagrangians in \p{LagrSO51}, \p{SUpc} and \p{385a} are superconformal. However, whereas
it is more or less clear that the superconformal actions should simultaneously enjoy the maximal
internal $R$-symmetry (appearing as the important
part of the relevant superconformal group) the converse is not always true: e.g.
the SO(4) invariant action
of  the multiplet ${\bf (4, 8, 4)}$, in the manifestly $N{=}8$ supersymmetric description
in the $N{=}8$ bi-harmonic
superspace \cite{BIS}, is not superconformally invariant, the same property should be shared
by the $N{=}4$ superfield counterpart  \p{484b} of this action. Thus the issue of
the superconformal invariance of the actions respecting maximal $R$-symmetries
requires a special analysis.

In the examples throughout the paper we constructed $N{=}8$ supersymmetric scalar potentials
by adding,  to the
sigma-model type superfield actions, the linear Fayet-Iliopoulos-type terms which
at the component level
amount to terms linear in the auxiliary fields. On the other hand, some $N{=}4$ multiplets
(e.g. ${\bf (3, 4, 1)}$ and ${\bf (4, 4, 0)}$ ) admit more general superfield potential
terms yielding in components,
apart from  the scalar potentials, also some minimal couplings to the background target
gauge fields
\cite{ikl,N4HSS,DI1}. It is interesting to inquire whether some of such $N{=}4$ generalized
potential terms can be completed to $N{=}8$ invariants.

The hierarchical relationships between different $N{=}8$ mechanics models with ${\bf 8}$
physical fermions were
established here at the off-shell component level. It would be desirable to see this hierarchy
structure at the
manifestly supersymmetric superfield level, at least in the $N{=}4$ superfield approach.
For the case of $N{=}4$ mechanics model it was shown in \cite{DI1,DI2} that the corresponding
hierarchical structure is directly related to
gauging, by non-propagating gauge $N{=}4, d{=}1$ multiplets, triholomorphic
(commuting with supersymmetry)
isometries of the ``parent'' $N{=}4$ mechanics model associated with the
$N{=}4$ ``root'' multiplet ${\bf (4, 8, 4)}$.
This superfield framework made it possible to reveal some new possibilities and features which
were not seen in the component approach.  It would be desirable to extend this gauging procedure
to the $N{=}8$ mechanics models as well. Though all homogeneous compact triholomorphic isometries
used in \cite{DI1,DI2} cease to be triholomorphic with respect to the full $N{=}8$ supersymmetry
(they do not commute with the hidden $N{=}4$ supersymmetry and become a part of the
relevant full $R$-symmetries), there still remain some non-compact isometries, the shift
and rescaling ones, which commute with both $N{=}4$ supersymmetries, manifest and hidden.
So these isometries can be appropriate candidates for gauging.
In the $N{=}4$ case, the gauging is most naturally performed in
the $N{=}4, d{=}1$ harmonic superspace \cite{N4HSS}.
Hence, as a prerequisite,  it would be helpful to reformulate the models studied
here within the $N{=}4$ harmonic superspace
framework.

The $N{=}8, d{=}1$ off-shell multiplets classified and studied in \cite{BIKL2,ABC} and in the present
paper are defined by linear constraints. On the other hand, in the $N{=}4$ case most
of analogous off-shell linear multiplets have
their nonlinear cousins \cite{N4HSS,ikl1,bks,ks,DI1,BurKS}. This fact raises the question whether
similar nonlinear counterparts exist for the linear $N{=}8$ multiplets and which new properties
(including the geometric ones) the relevant $N{=}8$ mechanics models could have. In particular,
it seems that the only possibility to obtain models with more general target geometries (not
reducing to the conformally flat ones) is to involve nonlinear multiplets. Until now not too
many nonlinear $N{=}8$ multiplets are known: in ref. \cite{BBKS} a nonlinear version of the
multiplet ${\bf (2, 8, 6)}$ was discussed and in refs.
\cite{BKM,EI}  a nonlinear version of the multiplet ${\bf (4, 8, 4)}$ was described.
It seems important to have the
full list of nonlinear $N{=}8$ multiplets and the $N{=}8$ mechanics models associated with
them, like the one
existing now for linear $N{=}8$ multiplets. The $N{=}4$ superfield approach could be appropriate
for this purpose.
One could try to pair into irreducible $N{=}8$ multiplets nonlinear $N{=}4$ multiplets or, say,
nonlinear multiplet with a linear one, and to study general conditions under which
the corresponding actions with manifest $N{=}4$ supersymmetry possess also the
hidden $N{=}4$ supersymmetry, as we did here for the linear multiplets.

At last, it is imperative to work out the appropriate manifestly $N{=}8$ supersymmetric superfield
approach which would allow one to construct and study $N{=}8$ mechanics models without imposing
any constraints
on the relevant superfield actions. One such approach exists, this is the $N{=}8$ bi-harmonic
superspace approach
\cite{BIS}, but it perfectly works only for the linear and nonlinear  ${\bf (4, 8, 4)}$  multiplets
and seems to be
not too helpful in the other cases. Another approach is just the $d{=}1$ reduction of
the standard harmonic
superspace \cite{harm}, also with a limited scope of applications to $N{=}8, d{=}1$ supermultiplets.
It is worth to seek for
some alternative more universal $N{=}8, d{=}1$ superfield approaches.

\section*{Acknowledgements}
The work of E.I. and A.S. was supported in part by the RFBR grant 06-02-16684,
the RFBR-DFG grant 06-02-04012-a, the grant DFG, project 436 RUS 113/669/0-3, the grant INTAS 05-7928
and a grant of Heisenberg-Landau program. They thank Institute of Theoretical
Physics of the Leibniz University of Hannover for the kind hospitality during the course
of this study. E.I. thanks Laboratoire de Physique, UMR5672 of CNRS and ENS Lyon, for
the warm hospitality at the final stage of this work.

\section*{Appendix}
We quote some useful identities for the $N{=}4$ superfields describing two
different ${\bf (3, 4, 1)}$ multiplets:
\bea
&&
D^l_a D^n_b v^{ik} = i\, \eps_{ab} \Big \{ \eps^{ni} \partial_t v^{lk} + \eps^{nk} \partial_tv^{li}
- \frac{1}{2}\,(\eps^{ni} \eps^{lk} + \eps^{nk} \eps^{li})\, F
\Big \}\,,\nn
&&
D^{ia} \omega^{kb} = 3i\, \eps^{ab} \partial_t v^{ik} - \frac{3i}{2}\,\eps^{ik} \eps^{ab} F\,, \nn
&&
D^{ia} F = -\frac{2}{3}\, \partial_t \omega^{ia}\,, \nn
&&
D^c_i D^d_k w^{ab}
= i \eps_{ik} \Big \{ \eps^{da} \partial_t w^{cb} + \eps^{db} \partial_t w^{ca}
- \frac{1}{2}\, (\eps^{da} \eps^{cb} + \eps^{db} \eps^{ca})\, H
\Big \}\,, \nn
&&
D^{ia} \beta^{kb} = 3i\, \eps^{ik} \partial_t w^{ab} -\frac{3i}{2}\,\eps^{ik} \eps^{ab} H\,, \nn
&&
D^{ia} H = -\frac{2}{3}\, \partial_t \beta^{ia}\,. \nonumber
\eea


\begin{thebibliography}{99}

\bibitem{Lima} R.~de Lima Rodrigues, \emph{``The quantum mechanics SUSY algebra:
an introductory review''}, {\tt [hep-th/0205017]}.
\bibitem{BMSV} R.~Britto-Pacumio, J.~Michelson, A.~Strominger, A.~Volovich,
\emph{``Lectures on superconformal quantum mechanics and multi-black hole moduli spaces''},
{\tt [hep-th/9911066]}.
\bibitem{AnS} A.V. Smilga, \emph{``Low dimensional sisters of Seiberg-Witten effective theory''},
In ``From fields to strings'', eds. M. Shifman et al, World Scientific, 2004, Vol. 1, pp. 523-558;
{\tt [hep-th/0403294]}.
\bibitem{BLY} D.~Bak, K.~Lee, P.~Yi, Phys. Rev. {\bf D62} (2000) 025009,
{\tt [hep-th/9912083]}.
\bibitem{AP} V.~Akulov, A.~Pashnev, Teor.~Mat.~Fiz. {\bf 56} (1983) 344.
\bibitem{FR} S.~Fubini, E.~Rabinovici, Nucl. Phys. {\bf B245} (1984) 17.
\bibitem{leva} E.~Ivanov, S.~Krivonos, V.~Leviant,
J.~Phys. A: Math. Gen. {\bf 22} (1989) 345.
\bibitem{ikl} E.~Ivanov, S.~Krivonos, O.~Lechtenfeld, JHEP {\bf 0303} (2003) 014,
{\tt [hep-th /0212303]}.
\bibitem{ikl1} E.~Ivanov, S.~Krivonos, O.~Lechtenfeld, Class.~Quantum Grav. {\bf 21} (2004) 1031,
{\tt [hep-th /0310299]}.
\bibitem{BIKL2} S.~Bellucci, E.~Ivanov, S.~Krivonos, O.~Lechtenfeld,
Nucl. Phys. {\bf B684} (2004) 321,
{\tt [hep-th/0312322]}.
\bibitem{Freed} D.~Z.~Freedman, P.~F.~Mende, Nucl. Phys. {\bf B344} (1990) 317.
\bibitem{GTow} G.~W.~Gibbons, P.~K.~Townsend, Phys. Lett. {\bf B454} (1999) 187,
{\tt [hep-th/9812034]}.
\bibitem{BGK} S.~Bellucci, A.~Galajinsky, S.~Krivonos, Phys. Rev. {\bf D68} (2003) 064010,
{\tt [hep-th/0304087]}.
\bibitem{GL+} A.~Galajinsky, O.~Lechtenfeld, K.~Polovnikov,  Phys. Lett. {\bf B643} (2006) 221,
{\tt [hep-th/0607215]}.
\bibitem{GR} S.~J.~Gates, Jr., L.~Rana,
\emph{``On Extended Supersymmetric Quantum Mechanics''},
Maryland Univ. Preprint UMDPP 93-24. Oct. 1994.
\bibitem{GR1} S.~J.~Gates, Jr., L.~Rana, Phys. Lett. {\bf B342} (1995) 132,
{\tt [hep-th/9410150]}.
\bibitem{PT} A.~Pashnev, F.~Toppan, J.~Math.~Phys. {\bf 42} (2001)) 5257,
{\tt [hep-th/0010135]}.
\bibitem{Geom} R.~A.~Coles, G.~Papadopoulos, Class. Quant. Grav. {\bf 7} (1990) 427; G.~Papadopoulos,
Class. Quantum Grav {\bf 7} (2000) 3715, {\tt [hep-th/0002007]}; C.~M.~Hull, \emph{``The geometry of
supersymmetric quantum mechanics''}, {\tt [hep-th/9911001]}.
\bibitem{GPS} G.~W.~Gibbons, G.~Papadopoulos, K.~S.~Stelle, Nucl. Phys. {\bf B508} (1997) 623,
{\tt [hep-th/9706207]}.
\bibitem{ABC} S.~Bellucci, E.~Ivanov, S.~Krivonos, O.~Lechtenfeld,
Nucl. Phys. {\bf B699} (2004) 226,
{\tt [hep-th/0406015]}.
\bibitem{DE} D.-E.~Diaconescu, R.~Entin, Phys. Rev. {\bf D56} (1997) 8045-8052,
{\tt [hep-th/9706059]}.
\bibitem{AnSm} A.~V.~Smilga, Nucl. Phys. {\bf B652} (2003) 93,  {\tt [hep-th/0209187]}.
\bibitem{BZu} B.~M.~Zupnik, Nucl. Phys. {\bf B554} (1999) 365, {\tt [hep-th/9902038]}.
\bibitem{ISm} E.~A.~Ivanov, A.~V.~Smilga, Nucl. Phys. {\bf B694} (2004) 473, {\tt [hep-th/0402041]}.
\bibitem{BKN} S.~Bellucci, S.~Krivonos, A.~Nersessian, Phys. Lett. {\bf B605} (2005) 181,
{\tt [hep-th/0410029]}.
\bibitem{BKNS1} S.~Bellucci, S.~Krivonos, A.~Nersessian, A.~Shcherbakov,
\emph{``2k-dimensional N=8 supersymmetric quantum mechanics''}, {\tt [hep-th/0410073]}.
\bibitem{BKS1} S.~Bellucci, S.~Krivonos, A.~Shcherbakov, Phys. Lett. {\bf B612} (2005) 283,
{\tt [hep-th/0502245]}.
\bibitem{harm}A. Galperin, E. Ivanov, V. Ogievetsky, E. Sokatchev,
Pis'ma ZhETF {\bf 40} (1984) 155 [JETP Lett. {\bf 40} (1984) 912];
A.S. Galperin, E.A. Ivanov, S. Kalitzin, V.I. Ogievetsky, E.S. Sokatchev,
Class. Quantum Grav. {\bf 1} (1984) 469.
\bibitem{book} A.S. Galperin, E.A. Ivanov, V.I. Ogievetsky, E.S. Sokatchev,
{\it Harmonic superspace}, Cambridge University Press, Cambridge 2001, 306 p.
\bibitem{BKSu} S.~Bellucci, S.~Krivonos, A.~Sutulin, Phys. Lett. {\bf B605} (2005) 406,
{\tt [hep-th/0410276]}.
\bibitem{BIS} S.~Bellucci, E.~Ivanov, A.~Sutulin, Nucl. Phys. {\bf B722} (2005) 297,
{\tt [hep-th/0504185]}.
\bibitem{Faux} M.~Faux, S.~J.~Gates, Jr., Phys. Rev. {\bf D71} (2005) 065002,
{\tt [hep-th/0408004]}.
\bibitem{root} S.~Bellucci, S.~Krivonos, A.~Marrani, E.~Orazi, Phys. Rev. {\bf D73} (2006) 025011,
{\tt [hep-th/0511249]}.
\bibitem{BBKS}S. Bellucci, A. Beylin, S. Krivonos, A. Shcherbakov, Phys. Lett. {\bf B633} (2006) 382,
{\tt [hep-th/0511054]}.
\bibitem{BKS} S.~Bellucci, S.~Krivonos, A.~Shcherbakov, Phys. Rev. {\bf D73} (2006) 085014,
{\tt [hep-th/0604056]}.
\bibitem{BKM} S.~Bellucci, S.~Krivonos, A.~Marrani, Phys. Rev. {\bf D74} (2006) 045005,
{\tt [hep-th/0605165]}.
\bibitem{EI} E.~Ivanov, Phys. Lett. {\bf B639} (2006) 579,
{\tt [hep-th/0605194]}.
\bibitem{bks} C.~Burdik, S.~Krivonos, A.~Shcherbakov, Czech. J. Phys. {\bf 55} (2005) 1357,
{\tt [hep-th/0508165]}.
\bibitem{ks} S.~Krivonos, A.~Shcherbakov, Phys. Lett. {\bf B637} (2006) 119,
{\tt [hep-th/0602113]}.
\bibitem{BurKS} C.~Burdik, S.~Krivonos, A.~Shcherbakov,
\emph{``Hyper-Kahler geometries and nonlinear supermultiplets''},
{\tt [hep-th/0610009]}.
\bibitem{Topp1} Z.~Kuznetsova, M.~Rojas, F.~Toppan, JHEP {\bf 0603} (2006) 098,
{\tt [hep-th/0511274]}; F.~Toppan, \emph{``Irreps and Off-shell Invariant Actions of
the N-extended Supersymmetric Quantum Mechanics''}, Fifth International Conference
on Mathematical Methods in Physics IC2006, April 24-28 2006, CBPF, Rio de Janeiro, Brazil,
{\tt [hep-th/0610180]}.
\bibitem{N4HSS} E.~Ivanov, O.~Lechtenfeld, JHEP {\bf 0309} (2003) 073, {\tt [hep-th/0307111]}.
\bibitem{DI1} F.~Delduc, E.~Ivanov, Nucl. Phys. {\bf B753} (2006) 211, {\tt [hep-th/0605211]}.
\bibitem{DI2}F.~Delduc, E.~Ivanov, Nucl. Phys. {\bf B770} (2007) 179, {\tt [hep-th/0611247]}.

\end{thebibliography}
\end{document}